\documentclass[twocolumn,nofootinbib,longbibliography,prb,superscriptaddress]{revtex4-2}

\usepackage{feynmp}
\usepackage{amsmath}
\usepackage{graphicx}
\usepackage{multirow}
\usepackage{latexsym}
\usepackage{textcomp}
\usepackage{verbatim}
\usepackage{color}
\usepackage{bm}
\usepackage{subfigure}
\usepackage{everysel}
\usepackage{keyval}
\usepackage{ragged2e}
\usepackage{dsfont}
\usepackage{amssymb}
\usepackage{enumitem}
\usepackage{pstricks}
\usepackage{mathtools}
\usepackage{braket}
\usepackage{slashed} 
\usepackage[colorlinks=true]{hyperref}

\usepackage{graphicx}
\usepackage{xcolor}
\usepackage{amsmath}
\usepackage{amsfonts}
\usepackage{bbm}

\usepackage{ulem}

\newcommand{\osum}{{%
    \setbox0\hbox{\circ}%
    \rlap{\hbox to \wd0{\hss\sum\hss}}\box0
}}

\def \sgn{\mbox{sgn\,}}
\newcommand{\tr}{{\textrm{tr}}}
\newcommand{\Tr}{{\textrm{Tr}}}

\newcommand{\cO}{{\mathcal{O}}}
\newcommand{\cT}{{\mathcal{T}}}
\newcommand{\cD}{{\mathcal{D}}}

\begin{document}

\title{Sign-Problem-Free Variant of Complex Sachdev-Ye-Kitaev Model}

\author{Byungmin Kang}
\thanks{Electronic Address: bkang119@kias.re.kr}
\affiliation{School of Physics, Korea Institute for Advanced Study, Seoul 02455, Korea}

\author{Junggi Yoon}
\thanks{Electronic Address: junggiyoon@gmail.com}
\affiliation{School of Physics, Korea Institute for Advanced Study, Seoul 02455, Korea}
\affiliation{Asia Pacific Center for Theoretical Physics, Postech, Pohang 37673, Korea}

\date{\today}

\begin{abstract} 
We construct a sign-problem free variant of the complex Sachdev-Ye-Kitaev (SYK) model which keeps all the essential properties of the SYK model, including the analytic solvability in the large-$N$ limit and being maximally chaotic. In addition to the number of complex fermions $N$, our model has an additional parameter $M$ controlling the number of terms in the Hamiltonian which we take $M \to \infty$ with keeping $M/N$ constant in the large-$N$ limit. While our model respects global $U(1)$ symmetry associated with the fermion number conservation, both the large-$N$ limit and the sign-problem free nature become explicit in the Majorana representation. We present a detailed analysis on our model, i.e., the random matrix classification based on the symmetry analysis, analytic approach, and the quantum Monte Carlo simulations. All these analysis show that our model exhibit a non-Fermi liquid (NFL) physics, a gapless fermionic system lying beyond the conventional Fermi liquid picture.

\end{abstract}

\maketitle

\section{Introduction}
The non-Fermi liquid (NFL) is a gapless fermionic matter that goes beyond the conventional Fermi liquid picture due to strong interactions. While understanding such strongly interacting systems is highly desirable as it contains interesting phases of matter such as the strange-metal phase in high-$T_c$ superconductivity~\cite{RevModPhys.73.797}, the infamous sign-problem~\cite{PhysRevLett.94.170201} dictates the fundamental difficulty of the problem. With  only a few exceptions~\cite{PhysRevLett.46.519, PhysRevD.24.2278, PhysRevB.24.4295, PhysRevLett.47.1628, PhysRevB.26.5033, PhysRevB.28.4059, PhysRevB.71.155115, sandvik2010computational, PhysRevB.91.241117, PhysRevLett.116.250601, PhysRevLett.117.267002, wei2017semigroup}, general tools to understanding the NFL physics remain elusive so far.

The Sachdev-Ye-Kitaev (SYK) model is an interacting fermionic matter showing a non-Fermi liquid and having non-trivial holographic dual~\cite{Sachdev:1992fk, kitaevfirsttalk,KitaevTalks}. The model is solvable via analytical methods~\cite{Polchinski:2016xgd, Jevicki:2016bwu, Maldacena:2016hyu, Jevicki:2016ito}, which is the large-$N$ limit with $N$ being the number of fermions. In the large-$N$ limit, the model features the emergent reparametrization symmetry at the strong coupling limit. Furthermore, the model is shown to be maximally quantum chaotic~\cite{Maldacena:2015waa, Maldacena:2016hyu} due to a pseudo-Nambu-Goldstone boson described by Schwarzian low energy effective action which is originated from the spontaneous and explicit breaking of the emergent reparametrization symmetry.

In the gravity system, black holes are shown to saturate the chaos bound~\cite{Roberts:2014isa,Shenker:2014cwa}. But, in the field theory, few models have been proven to saturate the chaos bound which makes the SYK model special. Since the SYK model does saturates the chaos bound as the black holes, it suggests that the SYK model could be holographically dual to a black hole. Indeed, it was shown~\cite{Maldacena:2016upp} that the two-dimensional Jackiw-Teitelboim~(JT) gravity describes the low energy sector of the SYK model.

Having all these nice features, the SYK model and its extensions have been extensively studied in the literature~\cite{Fu:2016vas, Cotler:2016fpe, Yoon:2017nig,Li:2017hdt,PhysRevB.95.134302, Garcia-Garcia:2017bkg, pan2020self, PhysRevB.103.195108, PhysRevResearch.2.033505, PhysRevResearch.2.043049, lantagne2020superconducting, haenel2021traversable}. But almost all of the works are based on either the exact diagonalization method or the large-$N$ approach, which do not go beyond the framework of the original studies~\cite{Sachdev:1992fk, kitaevfirsttalk, KitaevTalks}. In this regard, we additionally employ numerically unbiased quantum Monte Carlo (QMC) method to tackle the SYK physics, applying to a sign-problem free model we introduce in the following. The QMC simulations not only confirm the predictions from the analytical approach but also probe the physics beyond the regime of the analytical approach.

Before proceeding, we would like to comment on previous QMC studies on the SYK model~\cite{pan2020self, PhysRevB.103.195108}. The model presented in Refs.~\onlinecite{pan2020self, PhysRevB.103.195108} involves both fermions and bosons, whereas our model is written purely in terms of fermions and therefore we are able to simulate with larger number of fermions. Nonetheless, both the model in Refs.~\onlinecite{pan2020self, PhysRevB.103.195108} and our model show non-Fermi liquid (NFL) behaviors without any fine-tuning in the coupling constants.

\section{Model}
Our model is an interacting model of complex fermions in (0+1)-dimension, similar to the usual complex SYK model. In terms of fermion creation and annihilation operators, the Hamiltonian is given by
\begin{equation}
H = \frac{1}{2} \sum_{a=1}^M (\hat{S}_a)^2 = - \sum_{i,j,k,l=1}^N \frac{1}{2} \Big( \sum_{a=1}^M J_{a;ij} J_{a;kl} \Big) c_i^\dagger c_j c_k^\dagger c_l,
\label{eq:Ham-QMC-SYK}
\end{equation}
where $M$ controls the number of terms in the Hamiltonian and $\hat{S}_a = - i \sum_{j, k=1}^N J_{a;jk} c_j^\dagger c_k$ with $J_{a;jk}$ being real Gaussian random variables. Each $J_{a; jk}$ with $j > k$ is drawn from the Gaussian distribution $\mathcal{P}(x) \propto \exp \Big[ - \frac{NM x^2}{2J^2} \Big]$ and then we impose anti-symmetry condition $J_{a;kj} = -J_{a;jk}$. The anti-symmetricity on $J_a$ implies that $\hat{S}_a$ is a Hermitian operator. Note that unless $(i,j)$ equals $(k,l)$ or $(l,k)$, the coupling constants $- \frac{1}{2} \sum_{a=1}^M J_{a;ij} J_{a;kl}$ follow a distribution having zero mean and finite variance. This observation highlights the similarities between our model and the (particle-hole symmetrized) complex SYK model with real coupling constants, while the difference seems essential in making the model sign-problem-free. 

Though not obvious in the complex fermion representation, Eq.~\eqref{eq:Ham-QMC-SYK} is free from the negative-sign problem. The sign-problem-free nature becomes explicit in the Majorana representation~\cite{PhysRevB.91.241117, PhysRevLett.116.250601, PhysRevLett.117.267002} as Eq.~\eqref{eq:Ham-QMC-SYK} and its proper deformations are in the ``Majorana class''~\cite{PhysRevLett.117.267002} or satisfy ``Majorana reflection positivity''~\cite{PhysRevLett.116.250601}, where more details can be found in Appendix~\ref{App:DQMC}. In the following sections, we provide the random matrix classification by analyzing the symmetries of our model, the scaling solution from the large-$N$ limit with fixing the ratio $M/N$ constant, and the temporal Green's function from the quantum Monte Carlo simulations. Rather surprisingly, the analytic approach is also manageable in the Majorana representation, similar to proving that the model is sign-problem free.

\section{Symmetry Analysis}
\label{sec:symmetry}
Let us now discuss the symmetries of Eq.~\eqref{eq:Ham-QMC-SYK}\footnote{By symmetry, we always mean either unitary or anti-unitary operator that commutes with the second quantized Hamiltonian.}. An obvious one is $U(1)$ symmetry associated with the fermion number conservation $\mathcal{Q} = \sum_i c_i^\dagger c_i$.

The second one is the time-reversal symmetry $\mathcal{T}$
\begin{equation} \label{eq:TRS-complex}
\begin{cases}
    \mathcal{T} \big( i \big) \mathcal{T}^{-1} = -i \\
    \mathcal{T} \big( c_i \big) \mathcal{T}^{-1} = c_i \\
    \mathcal{T} \big( c_i^\dagger \big) \mathcal{T}^{-1} = c_i^\dagger 
\end{cases} ,
\end{equation}
which maps the vacuum state to the vacuum state $\mathcal{T} \vert {\rm vac} \rangle = \vert {\rm vac} \rangle$. Since $\mathcal{T}^2 \big( c_i \big) \mathcal{T}^{-2} = c_i$ and $\mathcal{T}^2 \big( c_i^\dagger \big) \mathcal{T}^{-2} = c_i^\dagger$, $\mathcal{T}$ squares to $1$ at the single-particle level. $\mathcal{T}$ squares to $1$ in the many-body Fock space as well. 

The final symmetry is the chiral symmetry $\mathcal{S}$\footnote{One can combine $\mathcal{T}$ and $\mathcal{S}$ to get the particle-hole operator, which is also a symmetry of Eq.~\eqref{eq:Ham-QMC-SYK}.}
\begin{equation} \label{eq:particle-hole}
\begin{cases}
    \mathcal{S} \big( i \big) \mathcal{S}^{-1} = -i \\
    \mathcal{S} \big( c_i \big) \mathcal{S}^{-1} = c_i^\dagger \\
    \mathcal{S} \big( c_i^\dagger \big) \mathcal{S}^{-1} = c_i 
\end{cases} ,
\end{equation}
which maps the vacuum state to the completely filled state $\mathcal{S} \vert {\rm vac} \rangle = \big( \prod_{i=1}^N c_i^\dagger \big) \vert {\rm vac} \rangle$. As expected for the chiral symmetry, $\mathcal{S}$ squares to $1$ at the single-particle level, i.e., $\mathcal{S}^2 \big( c_i \big) \mathcal{S}^{-2} = c_i$ and $\mathcal{S}^2 \big( c_i^\dagger \big) \mathcal{S}^{-2} = c_i^\dagger$ hold. But as an operator on the many-body Fock space, $\mathcal{S}^2 = +1$ if $N = 0, 1 \,\, ({\rm mod} \,4)$ and $\mathcal{S}^2 = -1$ if $N = 2, 3 \,\, ({\rm mod} \,4)$. Following the notation in Ref.~\onlinecite{PhysRevB.95.115150}, two distinct notions of the square of $\mathcal{S}$ are encoded as the single-particle phase $\gamma_{\rm sp} = 1$ and the many-body phase $\gamma_{\rm mb} = \pm 1$ where the sign of the latter depends on $N$. Note that the many-body phase $\gamma_{\rm mb} = -1$ cannot be gauged away~\cite{PhysRevB.95.115150}. 

Having presented the symmetries of the model, we now discuss the random matrix theory~(RMT) classification~\cite{PhysRevLett.110.084101, PhysRevB.95.115150, Li:2017hdt}. To this end, we first decompose the Fock space into different symmetry sectors of unitary symmetries. The Hamiltonian in each symmetry sector is then classified by the presence or absence of three operators: an anti-unitary commuting with the Hamiltonian ($T_+$), an anti-unitary anti-commuting with the Hamiltonian ($T_-$), and a unitary anti-commuting with the Hamiltonian ($\Lambda$), where we can make gauge choices in such a way that $(T_+)^2$ and $(T_-)^2$ are either $+1$ or $-1$ while $\Lambda^2 = +1$. Note that the presence of two operators implies the third operator via $\Lambda = T_+ T_-$. These operators in total give 10 different symmetry classes~\cite{zirnbauer1996riemannian, PhysRevB.55.1142}. 

In our case\footnote{We consider the case with $M>1$ in Eq.~\eqref{eq:Ham-QMC-SYK} since $M=1$ Hamiltonian is equal to a non-interacting Hamiltonian squared, which is not generic.}, we employ the $U(1)$ symmetry to decompose the Hilbert space into charge-$q$ sectors $(q=0,1,\cdots, N)$. The chiral symmetry $\mathcal{S}$ maps the charge-$q$ sector to the charge-$(N-q)$ sector, whereas the time-reversal symmetry $\mathcal{T}$ remains as an anti-unitary operator commuting with the Hamiltonian in each charge sector. This suggests that the charge sector with $q = \frac{N}{2}$ is distinguished from the ones with $q \ne \frac{N}{2}$. Using $T_+ = \mathcal{T}$, the Hamiltonian of the charge-$\big( q \ne \frac{N}{2} \big)$ sector is classified as the symmetry class AI and follows the Gaussian orthogonal ensemble~(GOE) level statistics. 

On the other hand, the classification of the charge-$(q=\frac{N}{2})$ sector depends on the parity of $\frac{N}{2}$, i.e., whether $N = 0 \,\, ({\rm mod } \,4)$ or $2 \,\, ({\rm mod } \,4)$. Note that the chiral symmetry $\mathcal{S}$ on the charge-$\big( q=\frac{N}{2} \big)$ sector becomes an anti-unitary commuting with the Hamiltonian, i.e., $U_S ( H^* ) U_S^\dagger = H$ where the chiral symmetry $\mathcal{S}$ and the Hamiltonian are represented as $U_S \mathcal{K}$ and $H$ on the charge-$\big( q=\frac{N}{2} \big)$ sector. Combining $\mathcal{T}$ and $\mathcal{S}$, we get a unitary symmetry $U_S$ acting on the charge-$\big( q=\frac{N}{2} \big)$ sector. We therefore have to consider an individual symmetry sector of $U_S$ for a complete RMT classification. 

When $N = 0 \,\, ({\rm mod } \,4)$, $(U_S)^2 = 1$, so we divide the charge-$\big( q=\frac{N}{2} \big)$ sector into the symmetric (S) sector satisfying $U_S = 1$ and the anti-symmetric (A) sector satisfying $U_S = -1$. In both sectors, we have $T_+ = \mathcal{T}$ which squares to $1$. Therefore, the Hamiltonian in the charge-$\big( q=\frac{N}{2} \big)$ sector is represented as $\left( \begin{smallmatrix} H_\mathbb{R} & 0 \\ 0 & H'_\mathbb{R} \end{smallmatrix} \right)$, where $H_\mathbb{R}$ and $H'_\mathbb{R}$ are two distinct real symmetric matrices following the GOE level statistics.

When $N = 2 \,\, ({\rm mod } \,4)$, $(U_S)^2 = -1$, so we divide the charge-$\big( q=\frac{N}{2} \big)$ sector into $U_S = +i$ sector ($+$ sector) and $U_S = -i$ sector ($-$ sector). Having divided into symmetry sectors, the time-reversal symmetry $\mathcal{T}$ no longer becomes the symmetry of each symmetry sector. Instead, $\mathcal{T}$ now maps $+$ sector to $-$ sector, and vice versa. Therefore, the Hamiltonian in the charge-$\big( q=\frac{N}{2} \big)$ sector is represented as $\left( \begin{smallmatrix} H_\mathbb{C} & 0 \\ 0 & H^*_\mathbb{C} \end{smallmatrix} \right)$, where $H_\mathbb{C}$ and $H^*_\mathbb{C}$ are Hamiltonians on $+$ and $-$ sector and are related by the complex conjugation. Since no further symmetries exist, $H_\mathbb{C}$ follows the Gaussian unitary ensemble (GUE) level statistics.

\section{Analytic Approach} \label{sec:analytics}
Similar to the usual SYK models, our model Eq.~\eqref{eq:Ham-QMC-SYK} is also solvable in the large-$N$ limit, where we take both $N, M \to \infty$ while $\frac{M}{N} = r$ is held fixed. Since the random coupling constant $J_{a;jk}$ is real and anti-symmetric in exchanging $j$ and $k$, our model will have emergent $O(N)$ symmetry after the disorder average. Therefore, the Majorana representation would be a more natural choice for the analytic approach.

Using the complex fermions, Majorana fermions $\chi_{j, \sigma}$ with $j=1,2,\cdots, N$ and $\sigma = \pm$ can be written as 
\begin{equation}
\begin{cases}
    \chi_{j, +} = \frac{1}{\sqrt{2}} \big( c_j + c_j^\dagger \big) \\
    \chi_{j, -} = \frac{1}{\sqrt{2} i} \big( c_j - c_j^\dagger \big)
\end{cases} .
\end{equation}
In terms of Majorana fermions, we have $\hat{S}_a = -\frac{i}{2}\sum_{j,k=1}^N J_{a;jk} \big( \chi_{j, +} \chi_{k, +} + \chi_{j, -} \chi_{k, -} \big)$. Before proceeding, we would like to comment on the symmetries of Eq.~\eqref{eq:Ham-QMC-SYK} relevant for the large-$N$ approach. First of all, we consider the following anti-unitary symmetry $\mathcal{T}_+$: $\mathcal{T}_+ \big( i \big) \mathcal{T}_+^{-1} = -i$ and $\mathcal{T}_+ \big( \chi_{j, \pm} \big) \mathcal{T}_+^{-1} = \chi_{j, \mp}$, which is nothing but the time reversal symmetry $\mathcal{T}$ Eq.~\eqref{eq:TRS-complex} up to a unitary transformation. And from the fermion number conservation, we have $SO(2) \cong U(1)$ symmetry: $\chi_{j, \sigma} \mapsto \sum_\rho O_{\sigma \rho} \chi_{j, \rho}$, where $O_{\sigma \rho}$ is an $SO(2)$ matrix. Finally by combining the time-reversal symmetry $\mathcal{T}_+$ and $SO(2)$ symmetry, we get $O(2)$ symmetry which will play a crucial role in constructing bi-local collective fields.

For the large-$N$ limit, we introduce a bosonic field $\phi_a$ for each $a \in \{1, \cdots, M\}$. Using the standard Hubbard-Stratonovich transformation, the Hamiltonian can be written as
\begin{equation} \label{eq: hamiltonian4}
	H={1\over 2}(\phi_a)^2 - i \hat{S}_a\phi_a ,
\end{equation}
where the corresponding Euclidean action is given by
\begin{equation} \label{eq: action with boson}
	S= \int d\tau \; \left[{1\over 2} \chi_{j,\sigma} \partial_\tau \chi_{j,\sigma} + {1\over 2}(\phi_a)^2-  i \hat{S}_a\phi_a\right] . 
\end{equation}
We then define bi-local collective fields $\Psi(\tau_1,\sigma_1;\tau_2,\sigma_2)$ and $\Phi(\tau_1,\tau_2)$ as
\begin{align}
\Psi(\tau_1,\sigma_1;\tau_2,\sigma_2)\equiv& {1\over N} \sum_{j=1}^N \chi_{j,\sigma_1}(\tau_1)\chi_{j,\sigma_2}(\tau_2) \label{eq:col-field1} \\
\Phi(\tau_1,\tau_2)\equiv& {1\over M} \sum_{j=1}^M \phi_a(\tau_1)\phi_a(\tau_2) .
\label{eq:col-field2} 
\end{align}
By integrating out the random coupling constants according to the Gaussian distribution $\mathcal{P}= \prod_{j>k} \exp \left[- { N M  [J_{a;jk}]^2\over  2 J^2}\right]$, the collective action in terms of the collective fields Eqs.~\eqref{eq:col-field1} and~\eqref{eq:col-field2} is given by
\begin{align}
S_\textrm{col}=&N\int d\tau \; \left[ -\frac{1}{2} \partial_\tau \Psi(\tau,\sigma;\tau',\sigma) \vert_{\tau' \rightarrow \tau}+  \frac{r}{2} \Phi(\tau,\tau)\right] \nonumber \\
&+ \frac{N}{2} \Tr \log \Psi - \frac{rN}{2} \tr \log \Phi \nonumber \\
&+{J^2N\over 4}  \sum_{\sigma_1,\sigma_2} \int d\tau_1d\tau_2\;   \Phi(\tau_1,\tau_2) \left[\Psi(\tau_1,\sigma_1;\tau_2,\sigma_2)\right]^2 \ .
\label{eq:col-action}
\end{align}
where $\Tr$ and $\tr$ denote the trace over $(\tau,\sigma)$ and $\tau$ space, respectively. The saddle point equations of the collective action leads to the Schwinger-Dyson equation for the two-point functions
\begin{widetext}
\begin{align}
	&-\partial_{\tau_1}\Psi(\tau_1,\sigma_1;\tau_2,\sigma_2)+\delta(\tau_{12})\delta_{\sigma_1 \sigma_2} 
	- J^2\sum_{\sigma_3} \int d\tau_3 \Phi(\tau_1,\tau_3)\Psi(\tau_1,\sigma_1;\tau_3,\sigma_3)\Psi(\tau_3,\sigma_3;\tau_2,\sigma_2)=0\ ,\\
	& r\Phi(\tau_1,\tau_2) - r\delta(\tau_{12}) + {1\over 2} J^2\sum_{\sigma_1,\sigma_3} \int d\tau_3\;   \left[\Psi(\tau_1,\sigma_1;\tau_3,\sigma_3)\right]^2\Phi(\tau_3,\tau_2)=0 \ .
\end{align}
\end{widetext}
In strong coupling limit $J\rightarrow \infty$, we consider the following scaling ansatz:
\begin{align} 
\Psi_0(\tau_1,\sigma_1;\tau_2,\sigma_2) &= \frac{b_{\sigma_1 \sigma_2} \sgn(\tau_{12})}{|\tau_{12}|^{2 \Delta_{\sigma_1\sigma_2}} } \nonumber  \\
\Phi_0(\tau_1,\tau_2) &= \frac{b_{\Phi}}{|\tau_{12}|^{ 2\Delta_\Phi}} ,\label{eq: scaling ansatz}
\end{align}
where $b_{\sigma_1 \sigma_2}$ and $b_\Phi$ are constants and $\Delta_{\sigma_1 \sigma_2}$ and $\Delta_{\Phi}$ are scaling exponents. Note that the classical solution is the (time-ordered) two point function of the fundamental fermion. 
\begin{align}
	\Psi_0(\tau_1,\sigma_1;\tau_2,\sigma_2)=\langle \cT [\chi_{j,\sigma_1}(\tau_1)\chi_{j,\sigma_2}(\tau_2)]\rangle .
\end{align}
While $\Psi_0(\tau_1,\pm;\tau_2,\pm)$ are anti-symmetric function in $\tau_{12}$, $\Psi_0(\tau_1,\pm;\tau_2,\mp)$ are not anti-symmetric for generic Hamiltonian. However, our model has $O(2)$ symmetry where the only $O(2)$ invariant two point function is given by
\begin{align}
{1\over 2}\left[\Psi_0(\tau_1,+;\tau_2,+)+ \Psi_0(\tau_1,-;\tau_2,-)\right]={b_\Psi\sgn(\tau_{12})\over |\tau_{12}|^{2\Delta_\Psi} }
\end{align}
and the time reversal symmetry implies that
\begin{align}
\Psi_0(\tau_1,+;\tau_2,+)=\Psi_0(\tau_1,-;\tau_2,-)={b_\Psi\sgn(\tau_{12})\over |\tau_{12}|^{2\Delta_\Psi} } .
\label{eq: pm symmetry two point function}
\end{align}

The reparametrization invariance of the action (with the kinetic term ignored) gives the relation between conformal dimensions:
\begin{equation} \label{eq:conformal-dimension-1}
	\Delta_\Phi + 2\Delta_{\Psi}=1 
\end{equation}
Using the scaling ansatz Eq.~\eqref{eq: scaling ansatz}, the Schwinger-Dyson equation reduces to
\begin{align} \label{eq:SD-scaling-step1}
	\delta_{\sigma_1\sigma_2} \,=\, & J^2  \sum_{\sigma_3} b_{\sigma_1\sigma_3}b_{\sigma_3\sigma_2}b_\Phi  c_A(\Delta_\Phi+\Delta_\Psi)c_A(\Delta_\Psi) \cr
	&\hspace{10mm}\times|w|^{4\Delta_\Psi+2\Delta_\Phi-2} \nonumber \\
	r \,=\, &  {J^2\over 2} \sum_{\sigma_1,\sigma_3} [b_{\sigma_1\sigma_3}]^2 b_\Phi  c_S(2\Delta_\Psi)c_S(\Delta_\Phi)\cr
	&\hspace{12mm}\times|w|^{4\Delta_\Psi+2\Delta_\Phi-2} ,
\end{align}
where $c_A(\Delta)$ and $c_S(\Delta)$ are defined by
\begin{align}
	c_A(\Delta)=&2i \cos(\pi \Delta)\Gamma(1-2\Delta) \nonumber \\
	c_S(\Delta)=&2\sin(\pi\Delta)\Gamma(1-2\Delta)
\end{align}
with $\Gamma$ being the Gamma function and we used the following identities
\begin{align}
	\int_{-\infty}^\infty d\tau \; e^{iw \tau} {\sgn(\tau)\over |\tau|^{2\Delta}} =& c_A(\Delta)\sgn(w)|w|^{2\Delta-1} \nonumber \\
	\int_{-\infty}^\infty d\tau \; e^{iw \tau} {1\over |\tau|^{2\Delta}} =& c_S(\Delta) |w|^{2\Delta-1} .
\end{align}
Further simplification is possible if we employ the identities
\begin{align}
	c_A(\Delta)c_A(1-\Delta)=&-{2\pi \cos\pi \Delta\over (1-2\Delta)\sin \pi \Delta} \nonumber \\
	c_S(\Delta)c_S(1-\Delta)=&-{2\pi \sin \pi \Delta\over (1-2\Delta)\cos \pi \Delta} ,
\end{align}
which reduces Eq.~\eqref{eq:SD-scaling-step1} into
\begin{align} \label{eq:SD-scaling-step2}
	& -J^2  b_{\Psi}^2 b_\Phi  {2\pi \cos \pi \Delta_\Psi \over (1-2\Delta_\Psi)\sin\pi \Delta_\Psi}=1 \nonumber \\
	&-J^2  b_{\Psi}^2 b_\Phi  {2\pi \sin \pi \Delta_{\Phi}\over (1-2\Delta_\Phi)\cos\pi \Delta_\Phi} =r .
\end{align}
Using the above equations and Eq.~\eqref{eq:conformal-dimension-1}, we get
\begin{align}
	{(1-4\Delta_\Psi)\over (1- 2\Delta_\Psi)\tan \pi \Delta_\Psi \tan 2\pi \Delta_\Psi}
	\,=\,{1\over  r}\label{eq: conformal dim eq q2} .
\end{align}
For a given $r$, the conformal dimension $\Delta_\Psi$ can be determined, and accordingly, the coefficient $b_{\Psi}^2b_\Phi$ is also fixed. In the range $\Delta_\Psi\in [0,{1\over 2}]$ where both $\Delta_\Psi$ and $\Delta_\phi$ are non-negative, there are two solutions of Eq.~\eqref{eq: conformal dim eq q2}. If we consider the limiting case $r \to 0$, which we explore more in Appendix~\ref{app:M=1}, one of them approaches 0 while the other goes to ${1\over 2}$. Hence, among two solutions of Eq.~\eqref{eq: conformal dim eq q2}, we take the smaller one for $\Delta_\Psi$.

We defer additional large-$N$ analysis to Appendix~\ref{app:analytics}. In Appendix~\ref{sec:higher-q-analytics}, we consider the generalization to $q$-body interactions, and in Appendix~\ref{sec: collective model 4 four point function}, we evaluate the Euclidean four point function and read off the conformal dimensions of operators which flow in the intermediate channel of four point function. Moreover, in Appendix~\ref{sec: otoc} we evaluate the real-time out-of-time-ordered correlator~(OTOC) and prove that the Lyapunov exponent $\lambda_L$ is ${2\pi \over \beta}$, implying that our model is maximally chaotic.

\section{Quantum Monte Carlo Simulations}
In this section, we present numerical results from quantum Monte Carlo simulations of our model Eq.~\eqref{eq:Ham-QMC-SYK}. In particular, we show that our model shows a non-Fermi liquid (NFL) behavior, confirming the analytic result discussed in Sec.~\ref{sec:analytics}. To this end, we employ the determinant quantum Monte Carlo (DQMC) method~\cite{PhysRevLett.46.519, PhysRevD.24.2278, PhysRevB.24.4295, PhysRevLett.47.1628, PhysRevB.26.5033, PhysRevB.28.4059}, which is a standard method in simulating interacting fermionic systems. We emphasize that the DQMC method provides a complementary approach to the exact diagonalization~(ED) and the analytic large-$N$ approach. This is due to the fact that the DQMC method can access larger system sizes than what ED can access and at the same time contains all the non-perturbative effects in $1/N$ which could be overlooked in the analytic approach. Moreover, DQMC simulations can compute physical observables which cannot be computed from the large-$N$ approach. Below, we compute the charge susceptibility by introducing the chemical potential term to the Hamiltonian. 

In our DQMC simulations, we always take the quenched average, i.e., first compute the physical observables in each disorder realization and then take disorder average. This is in contrast to the analytic approach where the annealed average is used by assuming the replica diagonal solution. It is known that taking a sufficient number of disorder averages is important in the random disordered systems~\cite{RevModPhys.58.801, PhysRevB.102.224204}. Here, we find $1000$ disorder averages are sufficient for our purpose, so we take $1000$ disorder averages in all of our simulations. For more details on the DQMC simulations including the error analysis, please refer to Appendix~\ref{App:DQMC} and~\ref{app:QMC-details}. We set $J=1$ in the remaining section.

\begin{figure}[t]
\centering\includegraphics[width=0.45\textwidth]{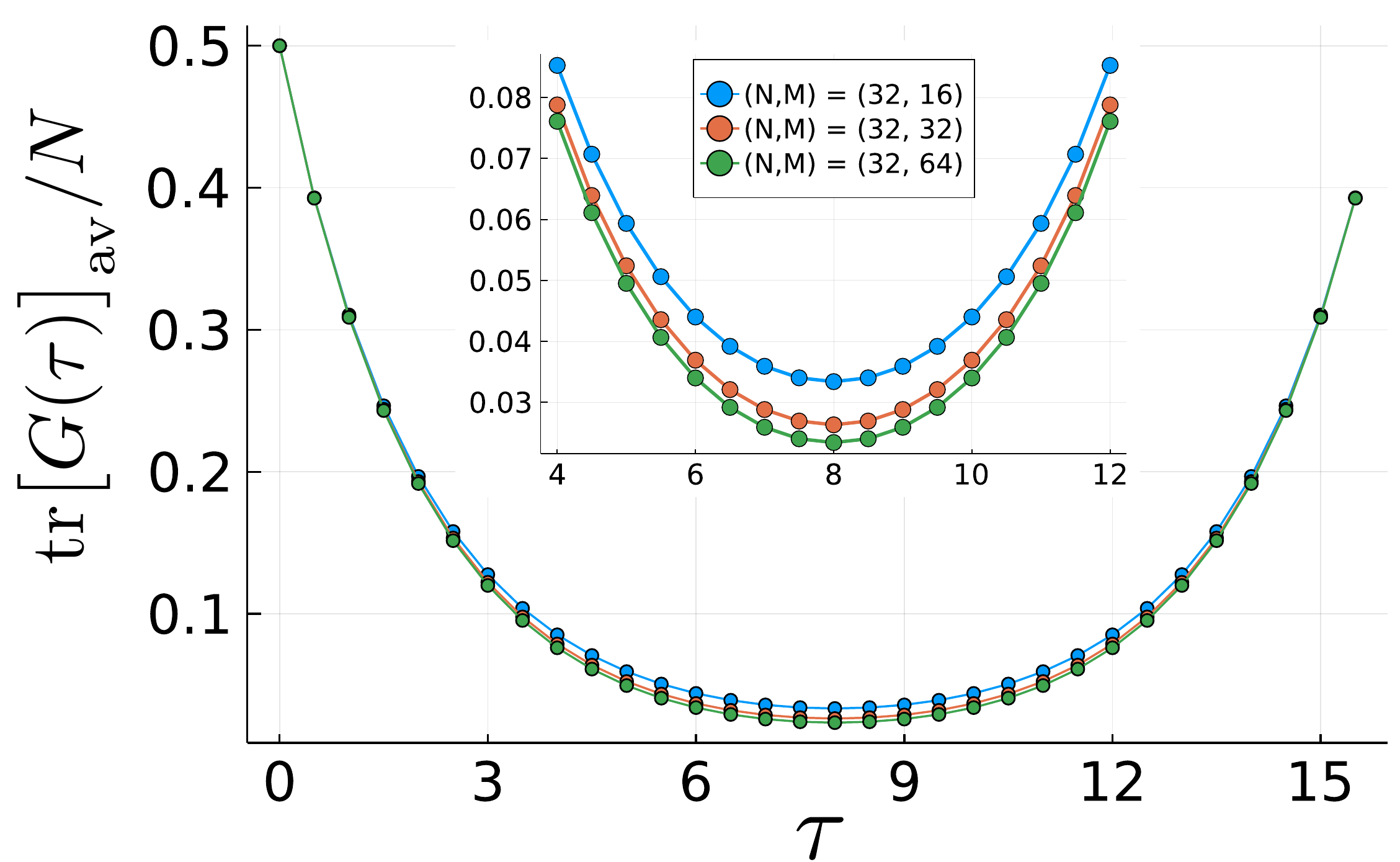}
\caption{Computation of $\tr \big[ G(\tau) \big]_{\rm av}/N$ from $1000$ disorder realizations with various $(N,M)$ at $\beta=16.0$ using the Trotter time step $\Delta \tau = 0.5$. (Inset) The same plot with range $\tau \in [4.0, 12.0]$. We see that the value $\tr \big[ G(0) \big]_{\rm av}/N$ decreases as $r = M/N$ increases. The statistical error from the Monte Carlo simulations is smaller than the size of the dot.}
\label{Fig:G-dis-av}
\end{figure}

\begin{figure}[t]
\centering\includegraphics[width=0.45\textwidth]{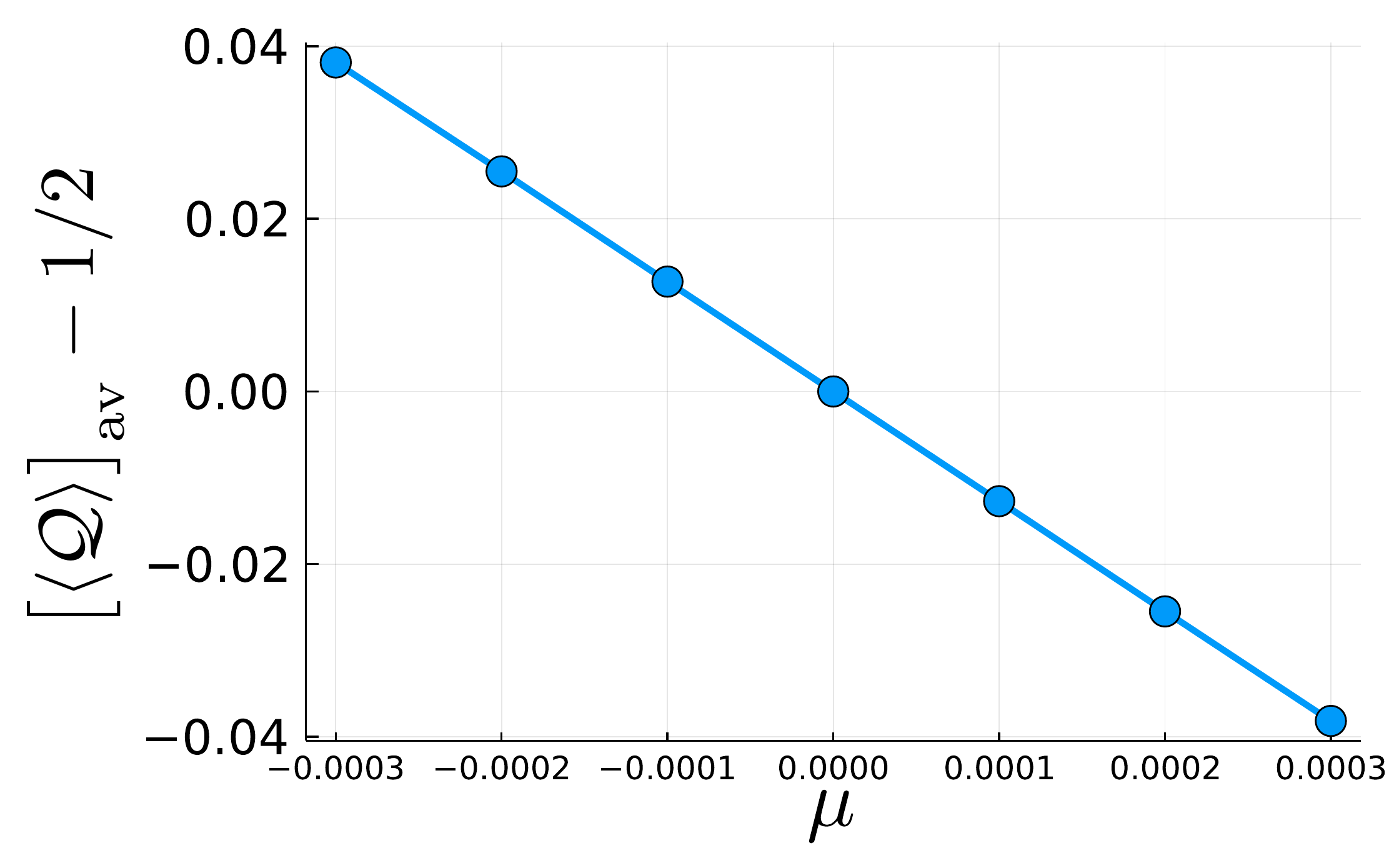}
\caption{Disorder averaged average charge $\big[ \langle \mathcal{Q} \rangle \big]_{\textrm{av}} - 1/2$ as a function of the chemical potential $\mu$ at $\beta=16.0$ using the Trotter time step $\Delta \tau = 0.5$. The statistical error from the Monte Carlo simulations error is smaller than the size of the dot. Since the charge changes as a function of $\mu$, it implies that the charge compressibility is non-zero and our system is gapless.}
\label{Fig:Q-vs-mu}
\end{figure}

Our main observable of interest is the disorder averaged temporal Green's function, which can easily be measured in the DQMC simulations. We first compare the disorder averaged Green's function for various $r=M/N$ in Fig.~\ref{Fig:G-dis-av}. From the figure, we notice that the Green's function decreases as the ratio $r$ increases. This is consistent with the observation made in the previous section that the conformal dimension $\Delta_\Psi$ increases as $r$ increases since the scaling regime ($\tau \approx \beta/2$) is dictated by the conformal dimension.

In Fig.~\ref{Fig:Q-vs-mu}, we compute the disorder averaged average charge $\big[ \langle \mathcal{Q} \rangle \big]_{\textrm{av}} - 1/2 = \frac{1}{N} \Big( \sum_{i=1}^N \big[ \langle c_i^\dagger c_i\rangle \big]_{\textrm{av}} \Big) - 1/2$ as a function of the chemical potential $\mu$, where $\textrm{av}$ denotes the disorder average and we add the chemical potential term $-\mu \sum_{i=1}^N c_i^\dagger c_i$ to our Hamiltonian Eq.~\eqref{eq:Ham-QMC-SYK}. Note that the average charge can be directly accessed from the Green's function via $\langle \mathcal{Q} \rangle = \tr \big[ G(0^+) \big]/N$. As can be seen from the figure, our model has non-zero charge compressibility $K = \frac{\partial \langle \mathcal{Q} \rangle}{\partial \mu}$, confirming the non-Fermi liquid behavior predicted from the analytical approach.

\section{Conclusions} 
In this work, we present a variant of the complex SYK model which is free from the sign-problem. While keeping all the analytic solvability in the usual SYK model, the sign-problem free nature allows one to understand a fully interacting system in an unbiased way. Using extensive DQMC simulations, we indeed confirm an exotic non-Fermi liquid behavior of the model as predicted from the analytical approach. 

Before conclude, we would like to mention interesting future directions. It is known that the SYK model have non-zero entanglement entropy~\cite{PhysRevB.97.245126, PhysRevD.100.041901, 10.21468/SciPostPhys.8.6.094, PhysRevResearch.2.033505} at $T=0$ if we take $N \to \infty$ first and then take the zero temperature limit. We can confirm the same feature in our model using both the large-$N$ limit and the DQMC simulations. This would reveal additional physical properties of our model. Also, we can study the phase transition in the eternal traversable wormhole suggested in Ref.~\onlinecite{maldacena2018eternal} directly via the quantum Monte Carlo by preparing two independent systems and couple them using random hopping terms. Moreover, our model could be used in studying exotic quantum phase transitions between the NFL phase and other phases such as superconducting phase and paramagnetic phases~\cite{PhysRevB.95.134302, PhysRevB.103.195108, lantagne2020superconducting}. Again, quantum Monte Carlo simulations would provide unbiased confirmations on the results from analytical approach and can go beyond the regime where the analytics apply.

Apart from the SYK models, it would be interesting to consider other classes of strongly interacting models using our approaches. In particular, fermionic tensor models without random coupling constant have been spotlighted recently because of the dominance of the melonic diagrams similar to the SYK model, and therefore those models are solvable at strong coupling limit~\cite{Witten:2016iux,Gurau:2016lzk,Klebanov:2016xxf,Narayan:2017qtw,Yoon:2017nig,Giombi:2018qgp}. Due to the similarity of the four point function, those tensor models are most likely maximally chaotic as in the SYK model. However, since the collective action for the tensor models in the large-$N$ limit is not known\footnote{A collective action for a subsector of the gauge invariant operators was found in Ref.~\onlinecite{deMelloKoch:2017bvv}.} unlike the SYK model, rigorous derivation for the saturation of chaos bound and investigation of phase transitions is not fully achieved\footnote{The rank-$3$ tensor model can be considered as $D$ numbers of $N\times N$ matrices, and its phase diagram was investigated in $N\gg D\gg 1$ limit in Refs.~\onlinecite{Azeyanagi:2017drg,Ferrari:2019ogc}.}. Since much less is known about tensor models, it would be highly desirable to use quantum Monte Carlo as an additional tool to study the tensor models.

\acknowledgements
We thank Masaki Tezuka and Yuki Suzuki for the helpful discussion. BK is supported by KIAS individual Grant PG069402 at Korea Institute for Advanced Study and the National Research Foundation of Korea (NRF) grant funded by the Korea government (MSIT) (No. 2020R1F1A1075569). JY was supported by KIAS individual Grant PG070102 at Korea Institute for Advanced Study and the National Research Foundation of Korea (NRF) grant funded by the Korea government (MSIT) (No. 2019R1F1A1045971). JY is supported by an appointment to the JRG Program at the APCTP through the Science and Technology Promotion Fund and Lottery Fund of the Korean Government. This is also supported by the Korean Local Governments - Gyeongsangbuk-do Province and Pohang City. This work was supported by the Center for Advanced Computation at Korea Institute for Advanced Study.

\appendix

\section{Determinant Quantum Monte Carlo}
\label{App:DQMC}
\subsection{Trotter decomposition and observables}
In this section, we show that our model Eq.~\eqref{eq:Ham-QMC-SYK} is free from the negative-sign problem. To demonstrate that our model can be simulated using the standard determinant quantum Monte Carlo (DQMC)~\cite{PhysRevLett.46.519, PhysRevD.24.2278, PhysRevB.24.4295, PhysRevLett.47.1628, PhysRevB.26.5033, PhysRevB.28.4059}, we consider the following partition function:
\begin{align}
Z &= \Tr \Big[ e^{-\beta H} \Big] = \Tr \Big[ \prod_{\tau=1}^{N_T} e^{-\Delta \tau H} \Big] \nonumber \\
&= \Tr \Big[ \prod_{\tau=1}^{N_T} e^{- \frac{\Delta}{2} \tau \sum_{a=1}^{M} \hat{S}_a^2} \Big] \approx \Tr \Big[ \prod_{\tau=1}^{N_T} \prod_{a=1}^M e^{- \frac{\Delta \tau}{2} \hat{S}_a^2} \Big] \nonumber \\
& \approx \Tr \Big[ \prod_{\tau=1}^{N_T} \prod_{a=1}^M \big( \frac{1}{2} \sum_{x_{a,\tau} = \pm 1} e^{i \sqrt{\Delta \tau} x_{a,\tau} \hat{S}_a} \big) \Big] \nonumber \\
&= \frac{1}{2^{N_T M}} \sum_{\{x_{a, \tau} = \pm 1 \}} \Tr \Big[ \prod_{\tau=1}^{N_T} \prod_{a=1}^M e^{ \sum_{i,j} [h_{a,\tau}]_{(i,j)} c_i^\dagger c_j} \Big] \nonumber \\
&= \frac{1}{2^{N_T M}} \sum_{\{x_{a, \tau} = \pm 1 \}} \det \Big[ \openone_{N} + \prod_{\tau, a} e^{[h_{a,\tau}]} \Big] ,
\label{eq:DQMC-decomp}
\end{align}
where $N_T$ is the number of Trotter steps, $\Delta \tau = \beta/N_{T}$, $[h_{a,\tau}]$ is an antisymmetric $N$-by-$N$ matrix with its $(i,j)$th entry being $\sqrt{\Delta \tau} x_{a,\tau} J_{a;ij}$. 

The main physical observable of interest is the temporal Green's function $G(\tau_1, \tau_2) \equiv \langle \mathcal{T}_\tau c_\alpha (\tau_1) c_\beta^\dagger (\tau_2) \rangle = G(\tau_1 - \tau_2)$ with the convention:
\begin{equation}
\langle \mathcal{T}_\tau c_\alpha (\tau_1) c_\beta^\dagger (\tau_2) \rangle  = 
\begin{cases}
\langle c_\alpha (\tau_1) c_\beta^\dagger (\tau_2) , \quad \,\,\,\, \tau_1 \ge \tau_2 \\
- \langle c_\beta^\dagger (\tau_2) c_\alpha (\tau_1) , \quad \tau_2 > \tau_1
\end{cases}
\end{equation}
The temporal Green's functions are evaluated using the standard methods in the DQMC~\cite{PhysRevD.24.2278, PhysRevB.24.4295}. 

\subsection{Proof of absence of negative sign-problem}
In the following, we use two different methods, the first one is based on the symmetry principle~\cite{PhysRevLett.117.267002} and the second one is based on the Majorana reflection positivity~\cite{PhysRevLett.116.250601}, to show that Eq.~\eqref{eq:DQMC-decomp} is free from the negative-sign problem. We then consider deformations preserving the Majorana reflection positivity, where the deformations often break several symmetries so that the remaining symmetries are not strong enough to fulfill the symmetry principle. 

Our goal is to show that the determinant appearing in the last equation of Eq.~\eqref{eq:DQMC-decomp} is non-negative for every assignment of $\{x_{a,\tau}\}$. To this end, we recall the Majorana representation:
\begin{equation}
\begin{cases}
c_j = \frac{1}{\sqrt{2}} \big( \chi_{j,+} + i \chi_{j,-} \big) \\
c_j^\dagger = \frac{1}{\sqrt{2}} \big( \chi_{j,+} - i \chi_{j,-} \big) ,
\end{cases}
\end{equation}
where $\chi_{j, +}$ and $\chi_{j,-}$ are Majorana fermions which square to $1/2$. Using the Majorana representation, we get
\begin{equation}
\sum_{j,k=1}^N [h_{a,\tau}]_{(j,k)} c_j^\dagger c_k = \frac{1}{2} \sum_{j,k=1}^N [h_{a,\tau}]_{(j,k)} \big( \chi_{j,+} \chi_{k,+} + \chi_{j,-} \chi_{k,-} \big) ,
\label{eq:DQMC-1ptl-ham}
\end{equation}
where we used the fact that $[h_{a,\tau}]$ is an antisymmetric matrix. It is immediate to show that Eq.~\eqref{eq:DQMC-1ptl-ham} respects three mutually anti-commuting anti-unitaries\footnote{Here, we denote that operators anti-commute if they anti-commute at the single-particle level.}, where the first two $\mathcal{T}_{\pm}$ are given by:
\begin{equation}
\begin{cases} \label{eq:majorana-T-pm}
\mathcal{T}_{\pm} \big( i \big) (\mathcal{T}_{\pm})^{-1} = -i \\
\mathcal{T}_{\pm} \big( \chi_{j, +} \big) (\mathcal{T}_{\pm})^{-1} = \pm \chi_{j, -} \\
\mathcal{T}_{\pm} \big( \chi_{j, -} \big) (\mathcal{T}_{\pm})^{-1} = \chi_{j, +}
\end{cases} ,
\end{equation} 
and the last one $\mathcal{T}'_+$ is given by:
\begin{equation}
\begin{cases}
\mathcal{T}'_+ \big( i \big) (\mathcal{T}'_+)^{-1} = -i \\
\mathcal{T}'_+ \big( \chi_{j, +} \big) (\mathcal{T}'_+)^{-1} = \chi_{j, +} \\
\mathcal{T}'_+ \big( \chi_{j, -} \big) (\mathcal{T}'_+)^{-1} = -\chi_{j, -}
\end{cases} .
\end{equation}
Note that $\mathcal{T}_+$ and $\mathcal{T}'_+$ ($\mathcal{T}_-$) square(s) to $1$ ($-1$) at the single-particle level. 

Using the symmetry principle~\cite{PhysRevLett.117.267002}, the presence of three mutually anti-commuting anti-unitaries completes the proof that Eq.~\eqref{eq:DQMC-1ptl-ham} is sign-problem-free. Majorana reflection positivity~\cite{PhysRevLett.116.250601} of Eq.~\eqref{eq:DQMC-1ptl-ham} is immediate, since it can be written as $\chi^T \left( \begin{smallmatrix} \frac{1}{2} [h_{a,\tau}] & 0 \\ 0 & \frac{1}{2} [h_{a,\tau}] \end{smallmatrix} \right) \chi$, where $\chi^T = (\chi_{1,+}, \cdots, \chi_{N_+}, \chi_{1,-}, \cdots, \chi_{N,-} )$ and $[h_{a,\tau}]$ is an anti-symmetric real matrix.

In the remaining section, we consider two deformations which preserve the Majorana reflection positivity. The first deformation is adding a chemical potential term $-\mu \sum_{j=1}^N c_j^\dagger c_j = - \mu \sum_{j=1}^N \big( i \chi_{j,+} \chi_{j,-} + \frac{1}{2} \big) = - \mu \frac{N}{2} + \chi^T \left( \begin{smallmatrix} 0 & -i \frac{\mu}{2} \textbf{1} \\ i \frac{\mu}{2} \textbf{1} & 0 \end{smallmatrix} \right) \chi$. The first deformation breaks $\mathcal{T}_-$ so that the remaining symmetries $\mathcal{T}_+$ and $\mathcal{T}'_+$ are not strong enough to apply the symmetry principle. But the deformation still respects the Majorana reflection positivity since $- \frac{\mu}{2} \textbf{1}$ is either positive or negative definite Hermitian matrix depending on the sign of $\mu$. Note that this deformation preserves the $U(1)$ symmetry.

The second deformation is a (random) mass deformation $\sum_{j,k=1}^N i K_{j,k} \big( \chi_{j,+} \chi_{k,+} - \chi_{j,-} \chi_{k,-} \big) = \chi^T \left( \begin{smallmatrix} i [K] & 0 \\ 0 & -i[K] \end{smallmatrix} \right) \chi$, where $[K]$ is an anti-symmetric (random) real matrix. This deformation breaks $\mathcal{T}'_+$ but preserves $\mathcal{T}_\pm$, hence the Hamiltonian after the deformation belongs to the ``Majorana-class''~\cite{PhysRevLett.117.267002} and is sign-problem-free. The Majorana reflection positivity also follows immediately and note that this deformation breaks the $U(1)$ symmetry.

\section{Details on DQMC Simulations} \label{app:QMC-details}
\subsection{Trotter error}
\begin{figure}[t]
\centering\includegraphics[width=0.4\textwidth]{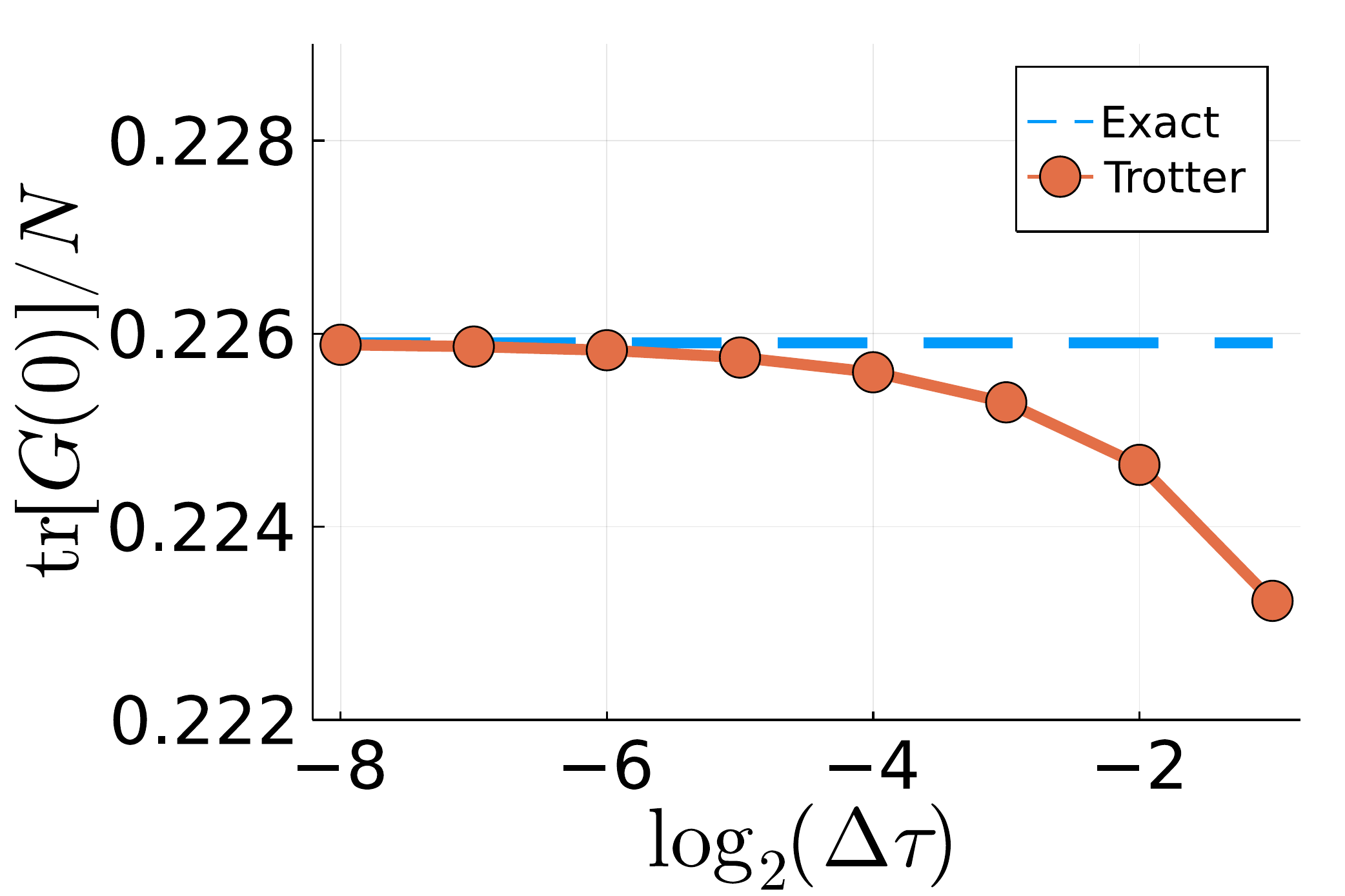}
\caption{Computation of $\tr \big[ G(0) \big]/N$ for a single disorder realization with $(N,M)= (8,8)$ and $\mu=1.0$ at $\beta=1.0$ using the exact partition function (denoted as ``Exact'') and the approximate partition function (denoted as ``Trotter'') in Eq.~\eqref{eq:DQMC-decomp}. Upon decreasing $\Delta \tau$, the result from the approximate partition function converges to the exact result.}
\label{Fig:ED-vs-Trotter}
\end{figure}
\begin{figure}[t]
\centering\includegraphics[width=0.4\textwidth]{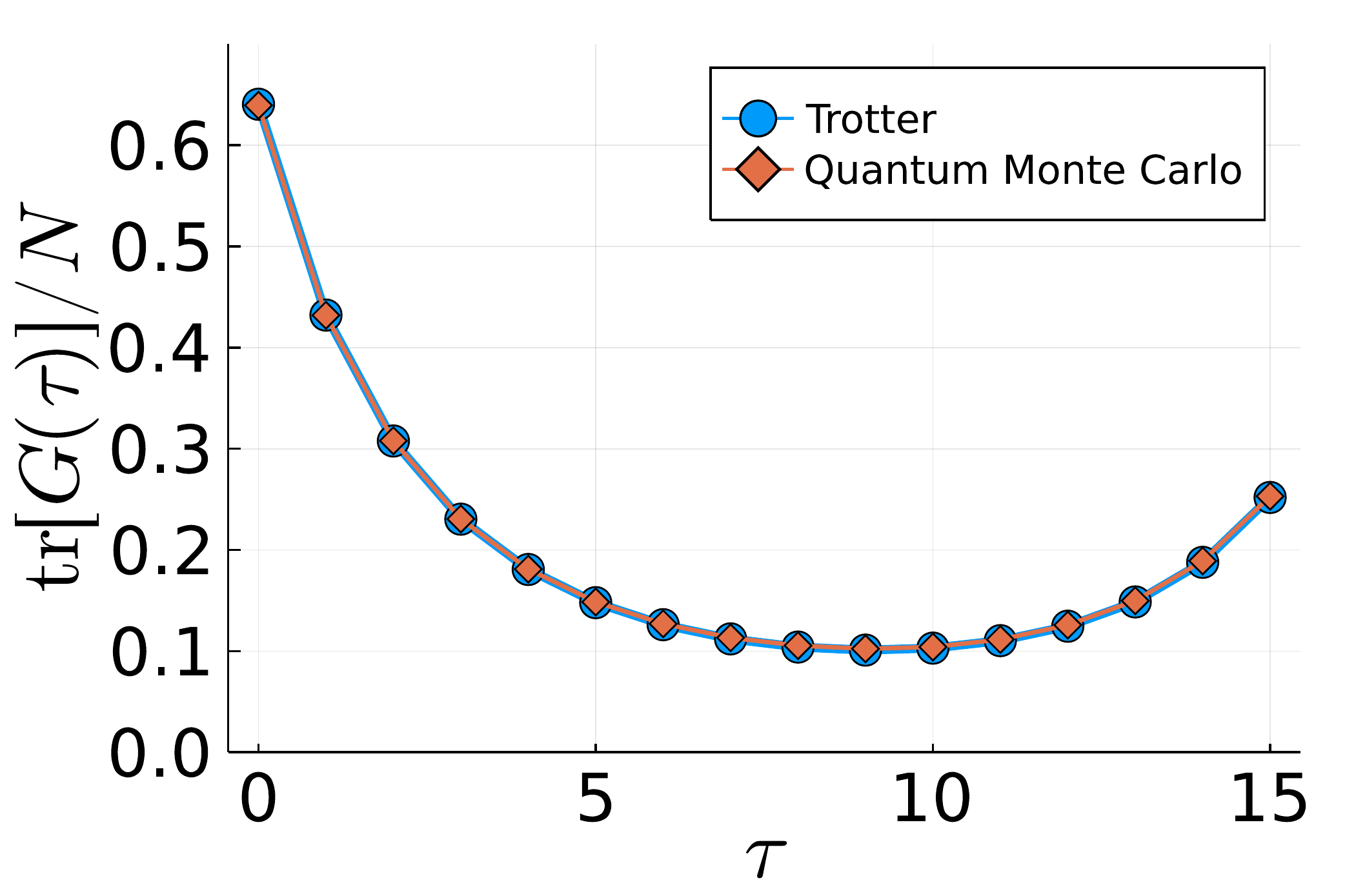}
\caption{Temporal Green's function computed from the Trotter approximated partition function Eq.~\eqref{eq:DQMC-decomp} for a single disorder realization with $(N,M) = (4,4)$ and $\mu = -0.01$ at $\beta = 16.0$, where we take $\Delta \tau = 1.0 $. Exact evaluation (denoted as ``Trotter'') and the computation using the DQMC simulation (denoted as ``Quantum Monte Carlo'') give identical results. In the DQMC simulation, we take $10^5$ number of measurements and the Monte Carlo error is smaller than the size of the dot.}
\label{Fig:Trotter-vs-QMC}
\end{figure}
In our DQMC simulations, it is crucial to take enough Trotter steps $N_T$ in order to ensure that the error from the Trotter discretization is vanishingly small. To demonstrate how the approximated partition function in Eq.~\eqref{eq:DQMC-decomp} converges to the exact partition function, we compute the Green's functions using the approximate partition function as a function of the Trotter time step $\Delta \tau = \beta/N_T$ and compare that with the Green's function using the exact partition function. In Fig.~\ref{Fig:ED-vs-Trotter}, we compute the trace of the Green's function $\tr \big[G(0) \big]/N$ for a single disorder realization of a small system $(N, M) = (8, 8)$ and $\mu=1.0$ at $\beta = 1.0$, where we include the chemical potential term $-\mu \sum_{i=1}^N c_i^\dagger c_i$ appropriately. Since the number of particles is rather small, the Green's functions are evaluated exactly using finite-size numerics. As expected, the Green's function from the approximated partition function indeed converges to the exact one as $\Delta \tau \to 0^+$. Furthermore, we confirm that our DQMC simulations correctly reproduce the results from the (Trotter approximated) partition function Eq.~\eqref{eq:DQMC-decomp}, which is demonstrated in Fig.~\ref{Fig:Trotter-vs-QMC}.

\subsection{Choice of Trotter time step}
\begin{figure}[t]
\centering\includegraphics[width=0.4\textwidth]{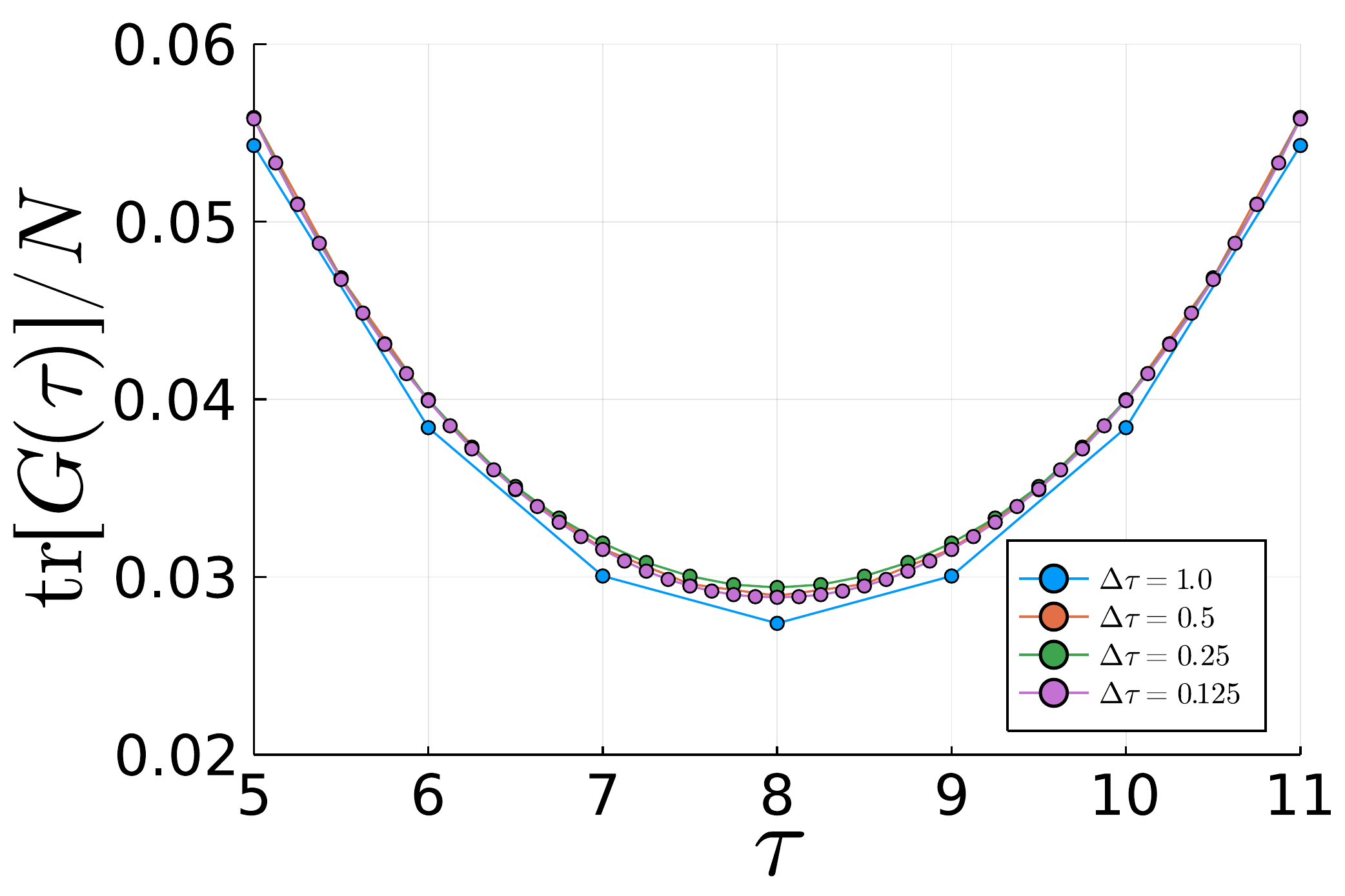}
\caption{The DQMC computation of the temporal Green's function as a function of the Trotter time step $\Delta \tau$ for a single disorder realization with $(N,M) = (32, 32)$, $\beta=16.0$, and $\mu=0.0$. We present the data near $\tau = \beta/2$ at which the differences between various $\Delta \tau$ become maximal. The statistical error from the Monte Carlo measurements is smaller than the size of the dot.}
\label{Fig:QMC-N_Trotter}
\end{figure}

While taking smaller Trotter time step $\Delta \tau$ decreases the Trotter error, the computation time increase accordingly. It is therefore important to take an optimal $\Delta \tau$ in the DQMC simulations. In Fig.~\ref{Fig:QMC-N_Trotter}, we compute $\tr \big[ G(\tau) \big]/N$ for various values of $\Delta \tau$ for a single disorder realization with $(N,M)=(32,32)$, $\mu=0.0$, and $\beta=16.0$. From the figure, we see that the results with $\Delta \tau \le 0.5$ lie almost on top of each other, where the difference becomes only noticeable near $\tau = \beta/2$. Guided by the figure, we take $\Delta \tau = 0.5$ as our optimal $\Delta \tau$, which is the value we used in all of our simulations presented in the main text.

\subsection{Disorder average and Monte Carlo error analysis} \label{app:MC-error}
In our DQMC simulations, we take the quenched average on physical observables, i.e., compute the physical observables for each disorder realization and then take the disorder average. To be specific, we consider the following the disorder averaged physical observable
\begin{equation} \label{eq:dis-av-obs}
[ \langle O \rangle ]_{\rm dis} = \frac{1}{N_{\rm dis}} \sum_{i=1}^{N_{\rm dis}} \langle O \rangle_{i} = \frac{1}{N_{\rm dis} N_m} \sum_{i=1}^{N_{\rm dis}} \sum_{j=1}^{N_m} O_{ij} ,
\end{equation}
where $O_{ij}$ is the $j$th measurement outcome of $i$th disorder realization and $\langle O \rangle_i$ is the Monte Carlo estimation for physical observable $O$ of the $i$th disorder realization. Note that the disorder average and the Monte Carlo measurement commute with each other for the observables of the form given by Eq.~\eqref{eq:dis-av-obs}, which includes the disorder averaged temporal Green's function, our main observable of interest. This suggests that we can compute the statistical error using $\langle O \rangle_{i =1, \cdots, N_{\rm dis}}$ only. To be specific, we think of each $\langle O \rangle_i$ as ``binning'' of $N_m$ samples $\{O_{ij}\}_{j=1,\cdots,N_m}$, and use the resulting uncorrelated $\langle O \rangle_i$ to estimate the statistical error using the standard bootstrap analysis.

\begin{figure}[t]
\centering\includegraphics[width=0.4\textwidth]{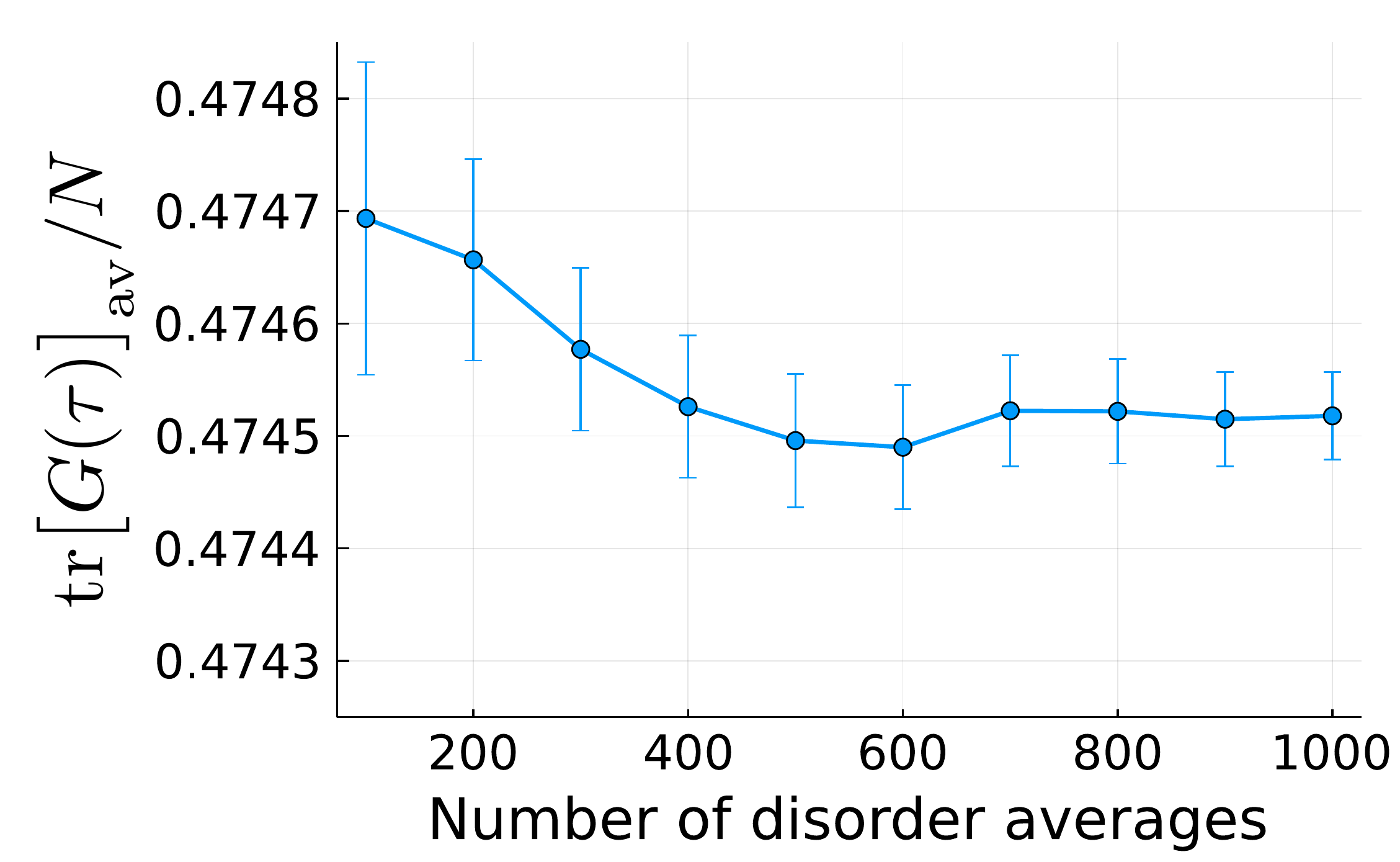}
\caption{Disorder average of the Green's function $\tr \big[ G(0) \big]/N$ as a function of the number of disorder averages for $(N,M) = (32,32)$, $\mu=0.0002$, and $\beta = 16.0$. For each disorder realization, we take $N_m = 10^4$ number of Monte Carlo measurements. The statistical error is estimated using the bootstrap error analysis described in Sec.~\ref{app:MC-error}.}
\label{Fig:QMC-err-vs-N_dis}
\end{figure}

In Fig.~\ref{Fig:QMC-err-vs-N_dis}, we present how the disorder averaged Green's function and its error change as a function of the number of disorder averages. Guided by the figure, we take $N_m=10^4$ Monte Carlo measurements for each disorder realization and take $N_{\rm dis} = 10^3$ disorder averages in all of the results presented in the main text. 

\section{$M=1$ Case}
\label{app:M=1}
In this section, we analyze $M=1$ case of our model Eq.~\eqref{eq:Ham-QMC-SYK}. Microscopically, $M=1$ case is different from other cases since the Hamiltonian of each disorder realization is given by a non-interacting Hamiltonian squared. When it comes to the large-$N$ limit, $M=1$ case would reveal the behavior of $r = M/N \to 0$ limit. 

The collective action for $M=1$ case can be written as
\begin{align}
S_\textrm{col}=&-\frac{N}{2}\int d\tau \;  \partial_\tau \Psi(\tau,\sigma;\tau',\sigma) \vert_{\tau' \rightarrow \tau}  + \frac{N}{2} \Tr \log \Psi \nonumber \\
&+{J^2N\over 4}  \sum_{\sigma_1,\sigma_2} \int d\tau_1d\tau_2\;   \phi(\tau_1)\phi(\tau_2) \left[\Psi(\tau_1,\sigma_1;\tau_2,\sigma_2)\right]^2\nonumber\\
& +  \frac{1}{2}\int d\tau\; [\phi(\tau)]^2 .
\end{align}
In the large-$N$ limit, we expect that the classical solution $\phi_{cl}(\tau)=\langle \phi(\tau) \rangle$ vanishes, and therefore the saddle point equation for $\Psi(\tau_1,\tau_2)$ is reduced to that of free case. This implies that the correct conformal dimension $\Delta_\Psi$ should approach to 0 in the limit $r \to 0$ as discussed in the main text.

\section{Exact Diagonalization Analysis}
In this section, we present numerical results obtained from the exact diagonalization (ED) approach. While the ED approach is limited to a small number of particles, it can access useful observables such as the level statistics and the spectral form factor (SFF) which cannot be accessed from other approaches. In the remaining section, we first focus on our main model Eq.~\eqref{eq:Ham-QMC-SYK} and analyze its level statistics and the SFF in each $U(1)$ charge sector. We then mix charge sectors by considering a mass deformation which breaks the $U(1)$ symmetry.

\subsection{Level statistics}
In Sec.~\ref{sec:symmetry}, we have classified each $U(1)$ charge sector of model Eq.~\eqref{eq:Ham-QMC-SYK} according to the random matrix theory (RMT). As a demonstration of our classification, we focus on the level statistic below. In each symmetry sector, we consider its spectrum $\{ E_1, E_2, \cdots \}$ in ascending order, i.e., $E_1 < E_2 < \cdots$. According to the ``Wigner-surmise''~\cite{PhysRevLett.110.084101}, the ratio $r_n = s_n/s_{n+1}$ of nearby energy spacing $s_n \equiv E_{n+1} - E_n$ follows the Wigner-Dyson level statistics:
\begin{equation} \label{eq:WD-surmise}
    p(r) = \frac{1}{Z} \frac{(r+r^2)^\beta}{(1+r+r^2)^{1+\frac{3}{2} \beta}},
\end{equation}
where $(\beta, Z) = \big(1, \frac{8}{27}\big)$ for the gaussian orthogonal ensemble (GOE), $(\beta, Z) = \big(2, \frac{4\pi}{81 \sqrt{3}}\big)$ for the gaussian unitary ensemble (GUE), and $(\beta, Z) = \big(4, \frac{4\pi}{729 \sqrt{3}}\big)$ for the gaussian symplectic ensemble (GSE). 

\begin{figure}[t]
\includegraphics[width=0.45\textwidth]{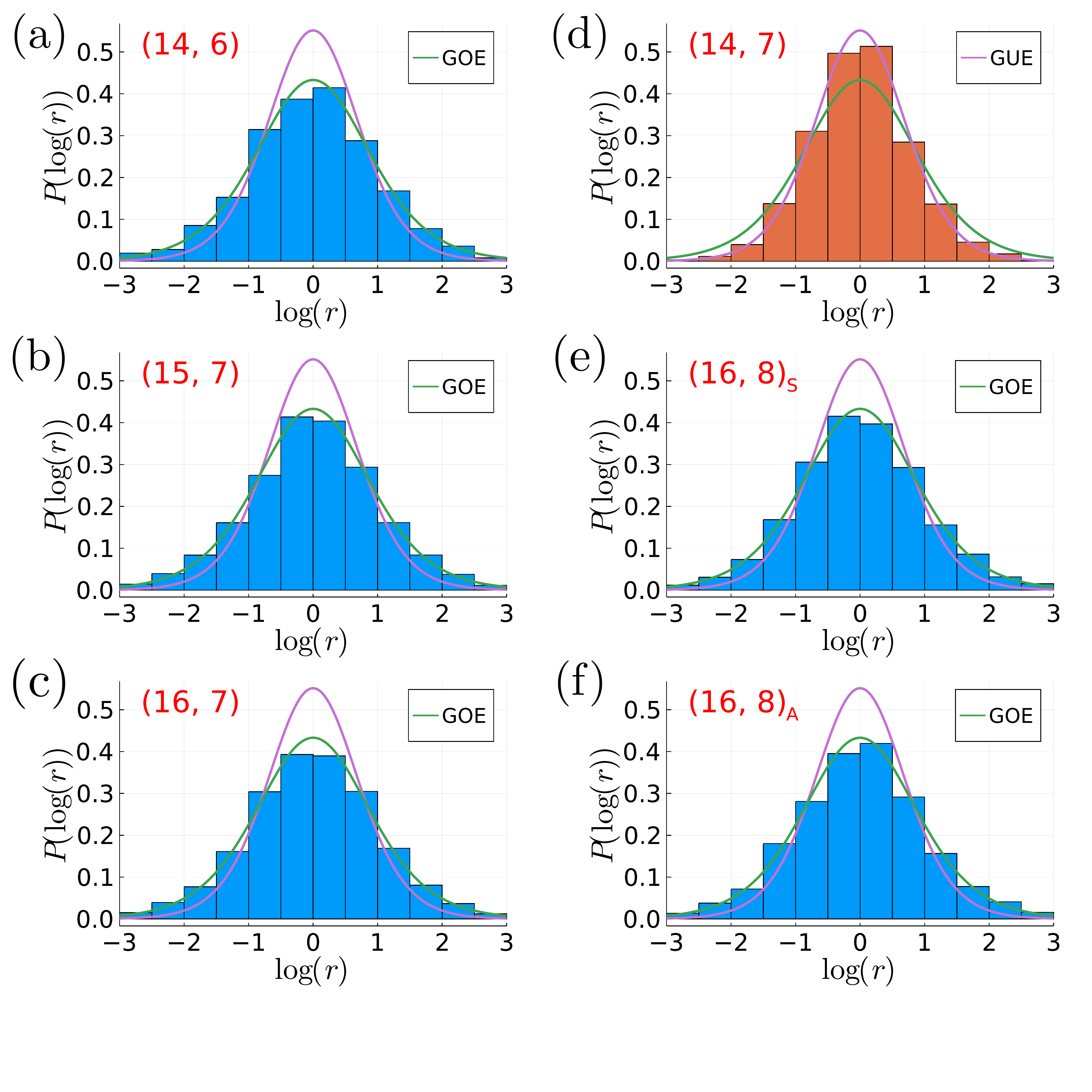}
\caption{Level statistics of a single disorder realization of the model Eq.~\eqref{eq:Ham-QMC-SYK} for various total number of orbitals $N$ and number of fermions $N_f$, where we set $M=N$ and denote $(N, N_f)$ in upper left corner in each plot. Two solid curves in each plot are from the exact level-statistic Eq.~\eqref{eq:WD-surmise} for GOE and GUE. When $(N, N_f)=(14, 7)$, $N = 2 \,\, ({\rm mod} \,4)$ and $N_f = N/2$ hold, so the corresponding $U(1)$ charge sector further splits into $+$ and $-$ sectors as discussed in Sec.~\ref{sec:symmetry}. Since the spectrum of two sectors are identical, we present the level statistics of one of the sector in (d). When $(N, N_f) = (16, 8)$, $N = 0 \,\, ({\rm mod} \,4)$ and $N_f = N/2$ hold, so we consider level statistics for (e) the symmetric (S) sector and (f) the anti-symmetric (A) sector, where $S$ sector and $A$ sector are defined in Sec.~\ref{sec:symmetry}. All cases except (d) follow the GOE level statistics while (d) follows the GUE level statistics, in accordance with the classification presented in Sec.~\ref{sec:symmetry}.}
\label{fig:level-statistics}
\end{figure}

In Fig.~\ref{fig:level-statistics}, we compute the level statistics for the different number of fermions and fillings. Our results indeed confirm the classification analyzed in Sec.~\ref{sec:symmetry}.  

\subsection{Spectral form factor}
The spectral form factor (SFF) is known to capture the chaotic behavior of the system~\cite{Cotler:2016fpe}. The SFF is defined as
\begin{equation} \label{eq:sff}
    g_c(t,  \beta) \, =\,  \bigg\langle \frac{|\sum_j e^{ - (\beta - i  t) E_j}|^2}{|\sum_j e^{- \beta E_j}|^2} \bigg\rangle_J ,
\end{equation}
where $\beta$ is the inverse temperature, $t$ is the real-time, and $\langle \; \cdot \;\rangle_J$ denotes the average over the random coupling constants. Also, its behavior has been reproduced by non-perturbative effects in 2D gravity~\cite{saad2018semiclassical, saad2019jt, Stanford:2019vob}. 

%
%
%
%

Traditionally, the quantum chaos, which is based on the random matrix theory via the BGS conjecture~\cite{Bohigas:1983er}, can be captured by the late time behavior of the SFF~\cite{Cotler:2016fpe}. Note on the other hand that the out-of-time-correlator (OTOC) measures the quantum chaos at early time before the scrambling time. The relation between two definitions of chaos, one from the random matrix theory and the other from the OTOC, was investigated in Ref.~\onlinecite{Nosaka:2018iat}. The SFF of the SYK model exhibits the same feature as that of the random matrix~\cite{Cotler:2016fpe}. 

\begin{figure}[t]
\includegraphics[width=0.45\textwidth]{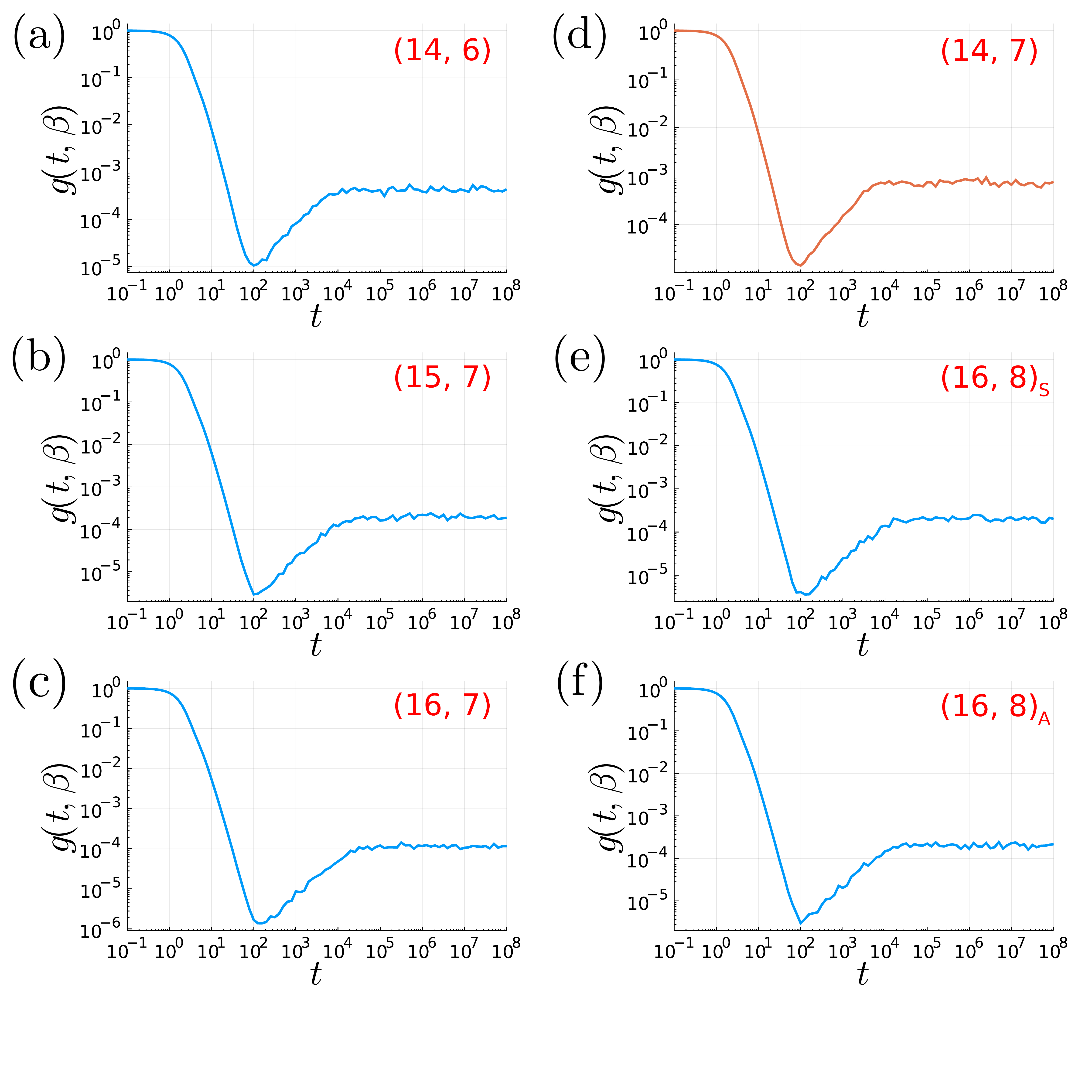}
\caption{Spectral form factor (SFF) of the model Eq.~\eqref{eq:Ham-QMC-SYK} using $100$ disorder disorder realizations at $\beta=1.0$ for various total number of orbitals $N$ and number of fermions $N_f$, where we set $M=N$ and denote $(N, N_f)$ in upper right corner in each plot. Please refer to the caption in Fig.~\ref{fig:level-statistics} for how we define the spectrum in the half-filled cases (d)-(f). The overall shapes of the SFF in all cases agree with that in the random matrix theory.}
\label{fig:spectral-form-factor}
\end{figure}

In Fig.~\ref{fig:spectral-form-factor}, we numerically evaluate the SFF Eq.~\eqref{eq:sff} for various charge sectors. We see a random matrix behavior, where the SFF shows an initial decay, followed by a dip, and then saturates to a plateau. If we evaluated the SFF for the whole system, the random matrix ensembles of each charge sector would be mixed. This leads to an oscillation of the SFF during the initial decay. But after the dip time, the behavior of the SFF of the whole sector exhibits the universal behavior as in the random matrix in spite of the mixture of ensembles, and it will be interesting to reproduce this universal behavior even from the two dimensional JT gravity together with $U(1)$ gauge field.

\subsection{Mass deformation}
\label{sec: mass deformation}
In this appendix, we consider the mass-deformations of the Hamiltonian Eq.~\eqref{eq:Ham-QMC-SYK} given by
\begin{align}
    H_{\text{\tiny mass}}\,\equiv \, \sum_{j, k=1}^N iK_{jk}\big(\chi_{j,+} \chi_{k,+} - \chi_{j,-} \chi_{k,-}  \big)\ .
\end{align}
where $K_{j,k}$ is a random coupling constant drawn from Gaussian distribution and anti-symmetric in $j, k$. Such a deformation makes the system less chaotic, and thus changes the characteristic of the system from maximally chaotic to a localized phase~\cite{PhysRevB.95.134302,Garcia-Garcia:2017bkg,Nosaka:2018iat}. When it comes to the holography, the mass deformation would provide an interesting phase diagram of the black holes. However, it is difficult to study this phase transition via the analytic method because the scaling ansatz for the two point function is not valid with the deformation. In the following, we discuss the random matrix behavior of the mass deformed Hamiltonian by analyzing symmetries.

In terms of the complex fermion, the mass deformation is given by $H_{\text{\tiny mass}} = \sum_{j,k=1}^N i K_{jk} \big(c_j c_k - c_k^\dagger c_j^\dagger)$. Hence the mass deformation breaks the $U(1)$ symmetry, the time-reversal symmetry $\mathcal{T}$ Eq.~\eqref{eq:TRS-complex}, and the chiral symmetry $\mathcal{S}$ Eq.~\eqref{eq:particle-hole}, but respects the particle-hole symmetry $\mathcal{T} \mathcal{S}$. Since the total fermion number is no longer conserved, it is convenient to introduce the fermion parity operator $P$:
\begin{equation} \label{eq:parity}
P = (-2i)^N \chi_{1,+} \chi_{1, -} \cdots \chi_{N,+} \chi_{N,-},
\end{equation}
which squares to one $P^2 = 1$ and commutes with the Hamiltonian. We then introduce two unitary operators
\begin{equation}
\begin{cases} \label{eq:unitary-Cpm}
P_+ = 2^{N/2} \chi_{1,+} \chi_{2,+} \cdots \chi_{N,+} \\
P_- = 2^{N/2} \chi_{1,-} \chi_{2,-} \cdots \chi_{N,-}
\end{cases} ,
\end{equation}
where we have
\begin{equation}
(P_+)^2 = (P_-)^2 = 
\begin{cases}
+1 \quad \textrm{if } N = 0, 1 \,\, ({\rm mod} \,4) \\
-1 \quad \textrm{if } N = 2, 3 \,\, ({\rm mod} \,4)
\end{cases}
\end{equation}
and
\begin{equation}
\begin{cases}
P_a \big( \chi_{j, a} \big) (P_a)^{-1} = (-1)^{N-1} \chi_{j, a} \\
P_a \big( \chi_{j, -a} \big) (P_a)^{-1} = (-1)^N \chi_{j, -a}
\end{cases} 
\end{equation}
with $a = \pm$. In fact, $P_+$ preciesly implements the particle-hole symmetry $\mathcal{T} \mathcal{S}$. In our convention, the complex conjugation $\mathcal{K}$ acts as $\mathcal{K} \big( \chi_{j, \pm} \big) \mathcal{K}^{-1} = \pm \chi_{j, \pm}$. Finally, we consider the following unitary operator
\begin{equation}
U_f = e^{i \frac{\pi}{4} N} \Big( \frac{1 - 2 \chi_{1,+} \chi_{1,-}}{\sqrt{2}} \Big) \cdots \Big( \frac{1 - 2 \chi_{N,+} \chi_{N,-}}{\sqrt{2}} \Big)
\end{equation}
which satisfies 
\begin{equation}
\begin{cases}
U_f \big( \chi_{j,+} \big) (U_f)^{-1} = \chi_{j,-} \\
U_f \big( \chi_{j,-} \big) (U_f)^{-1} = -\chi_{j,+}
\end{cases} .
\end{equation}
Note that $(U_f)^2$ equals the fermion parity operator $P$.

Now, two anti-unitary operators $\mathcal{T}_\pm$ Eq.~\eqref{eq:majorana-T-pm} are realized in the many-body Fock space as
\begin{equation}
\mathcal{T}_+ = U_f \mathcal{K} \\
\end{equation}
and
\begin{equation}
\mathcal{T}_- = 
\begin{cases}
P_+ U_f \mathcal{K} \qquad \qquad \,\,\, \textrm{if $N$ is odd} \\
(-i)^N P_- U_f \mathcal{K} \quad \quad \textrm{if $N$ is even}
\end{cases} ,
\end{equation}
where the phase factors are chosen such that $\mathcal{T}_+ \vert \textrm{vac} \rangle = \vert \textrm{vac} \rangle$ and $\mathcal{T}_- \vert \textrm{vac} \rangle = c_1^\dagger c_2^\dagger \cdots c_N^\dagger \vert \textrm{vac} \rangle$ hold. In the many-body Fock space, 
\begin{equation}
(\mathcal{T}_+)^2 = +1 .
\end{equation}

Two anti-unitary operators $\mathcal{T}_+$ and $\mathcal{T}_-$, equivalently an anti-unitary $\mathcal{T}_+$ and a unitary $P_+$, commute with the mass deformed Hamiltonian. Note also that the Hamiltonian is block diagonal in the parity even ($P=+1$) and the parity odd ($P=-1$) sectors. While $\mathcal{T}_+$ remains as an anti-unitary operator commuting with the Hamiltonian and squares to $+1$ in each parity sector, $P_+$ (anti-)commutes with $P$ when $N$ is even (odd). So for even $N$, one can further block diagonalize the Hamiltonian using $P_+$ eigenspaces. After a straightforward computation, one gets
\begin{equation}
P_+ \mathcal{T}_+ = i^N P \mathcal{T}_+ P_+ 
\end{equation}
which can be used to identify whether $\mathcal{T}_+$ changes $P_+$ eigenvalue or not.

When $N$ is even, We divide even $N$ case into two subcases $N = 0\,\, ({\rm mod} \,4)$ and $N = 2\,\, ({\rm mod} \,4)$. When $N = 0\,\, ({\rm mod} \,4)$, $(P_+)^2 = +1$, so its eigenvalues are $+1$ and $-1$. So on $P=1$ ($P=-1$) subspace $\mathcal{T}_+$ (anti-)commutes, and therefore $\mathcal{T}_+$ preserves (changes) the $P_+$ eigenvalue. When $N = 2\,\, ({\rm mod} \,4)$, $(P_+)^2 = -1$, so its eigenvalues are $+i$ and $-i$. So on $P=1$ ($P=-1$) subspace $\mathcal{T}_+$ anti-commutes (commutes), but due to anti-unitarity of $\mathcal{T}_+$, $\mathcal{T}_+$ preserves (changes) the $P_+$ eigenvalue.

To summarize, we get the following random matrix classification. Let us first represent the Hamiltonian using a Fock space basis as
\begin{equation}
H = \left(
\begin{array}{cc}
H_+ & 0 \\
0 & H_-
\end{array}
\right),
\end{equation}
where $H_+$ ($H_-$) is the Hamiltonian of the parity even (odd) sector. 

When $N$ is odd, $H_+$ and $H_-$ are real symmetric matrices related to each other via $P_+$ and thus have the same spectrum. Since $\mathcal{T}_+$ commutes with $H_+$, $H_+$ follows GOE level statistics.

When $N$ is even, $H_+ = \left( \begin{smallmatrix} H_\mathbb{R} & 0 \\ 0 & H'_\mathbb{R} \end{smallmatrix} \right)$ and $H_- = \left( \begin{smallmatrix} H_\mathbb{C} & 0 \\ 0 & H'_\mathbb{C} \end{smallmatrix} \right)$, where $H_\mathbb{R}$ and $H'_\mathbb{R}$ are distinct real symmetric matrices commuting with $\mathcal{T}_+$ and thus follow GOE level statistics. Also, $H_\mathbb{C}$ and $H'_\mathbb{C}$ are two complex Hermitian matrices following GUE level statistics. $H_\mathbb{C}$ and $H'_\mathbb{C}$ are related by $\mathcal{T}_+$ and thus have the same spectrum. Note that $H_\mathbb{R}$ and $H_\mathbb{C}$ ($H'_\mathbb{R}$ and $H'_\mathbb{C}$) are Hamiltonians on $P_+ = +1$ ($P_+ = -1$) subspace when $N = 0\,\, ({\rm mod} \,4)$ and on $P_+ = +i$ ($P_+ = -i$) subspace when $N = 2\,\, ({\rm mod} \,4)$. 

\section{Additional Details on Analytic Approach}
\label{app:analytics}
\subsection{Generalizations to higher order interaction}
\label{sec:higher-q-analytics}
One can generalize the Hamiltonian Eq.~\eqref{eq: action with boson} with higher order interaction. Defining
\begin{align}
    \hat{S}_a\equiv - i^{q\over 2}\!\! \sum^N_{j_1>\cdots>j_q}\! J_{a;j_1\cdots j_q} (  \chi_{j_1,+}  \cdots  \chi_{j_q,+}  +  \chi_{j_1,-}  \cdots  \chi_{j_q,-}  )\ ,
\end{align}
the action can be written as follows
\begin{align}
S=& \int d\tau \; \Big[{1\over 2} \chi_{j,\sigma} \partial_\tau \chi_{j,\sigma} + {1\over 2}(\phi_a)^2 -   i \phi_a  \hat{S}_a \Big]\ ,
\end{align}
where the random coupling constant $J_{a;j_1\cdots j_q}$ is drawn from the Gaussian distribution
\begin{equation}
	\mathcal{P}=\prod_{j_1>\cdots> j_q}\exp\left[ - {  N^{q-1}M [J_{a;j_1\cdots j_q}]^2 \over 2(q-1)! J^2 } \right]\ .
\end{equation}
After the disorder average, the collective action~\cite{Jevicki:2016bwu} is found to be
\begin{align}
&S_{col}=N\int d\tau \; \left[ -{1\over 2} \left.\partial_\tau \Psi(\tau,\sigma;\tau',\sigma)\right|_{\tau'\rightarrow \tau}+  {r\over 2} \Phi(\tau,\tau)\right] \nonumber \\
&+{N\over 2} \Tr \log \Psi - {rN\over 2} \tr \log \Phi \nonumber \\
&+{J^2N\over 2q}  \int d\tau_1d\tau_2\; \Phi(\tau_1,\tau_2)\cr 
&\hspace{10mm}\times\Big(   \left[\Psi(\tau_1,+;\tau_2,+)\right]^q + \left[\Psi(\tau_1,-;\tau_2,-)\right]^q \cr
&\hspace{14mm} + \left[\Psi(\tau_1,+;\tau_2,-)\right]^q+\left[\Psi(\tau_1,-;\tau_2,+)\right]^q\Big) .\label{eq: collective action q}
\end{align}
In the same way as in $q=2$ case, the scaling ansatz~\eqref{eq: scaling ansatz} for the classical solution of the bi-locals gives the equations for the conformal dimensions $\Delta_\Psi, \Delta_\Phi$ and the coefficients $b_{\Psi},\, b_\Phi$.
\begin{align}
	\Delta_\Phi+ q\Delta_\Psi\,=\,&1\label{eq: conformal dim eq q 1}\\
	 -J^2 b_{\Psi}^q b_\Phi  {2\pi \cos \pi \Delta_\Psi \over (1-2\Delta_\Psi)\sin\pi \Delta_\Psi}\,=\,&1\\
	- J^2 b_{\Psi}^q b_\Phi  {2\pi \sin \pi \Delta_{\Phi}\over (1-2\Delta_\Phi)\cos\pi \Delta_\Phi} \,=\,& {qr\over2}\ ,
\end{align}
and we have an equation determining the conformal dimension $\Delta_\Psi$
\begin{align}
	&{ 1-2q\Delta_\Psi\over (1- 2\Delta_\Psi)\tan \pi \Delta_\Psi \tan q\pi \Delta_\Psi }={2
	\over qr}\ .\label{eq: conformal dim eq q 2}
\end{align}
Demanding that the conformal dimensions are positive, we obtain the range of the $\Delta_\Psi$.
\begin{align}
    0<\Delta_\Psi<{1\over q}\ .
\end{align}
In this range of $\Delta_\Psi$, there exist two solutions of Eq.~\eqref{eq: conformal dim eq q 2}.

%
%
%

\begin{widetext}
\subsection{Four Point Function}
\label{sec: collective model 4 four point function}

We will study Euclidean four point functions $F_{\eta\eta}$, $F_{\phi,\eta}$ and $F_{\phi\phi}$ defined by
\begin{align}
	F_{\eta\eta}(\tau_1,\sigma_1;\tau_2,\sigma_2;\tau_3,\sigma_3;\tau_4,\sigma_4)\equiv & {1\over N^2} \langle \chi_{j,\sigma_1}(\tau_1)\chi_{j,\sigma_2}(\tau_2)\chi_{k,\sigma_3}(\tau_3)\chi_{k,\sigma_4}(\tau_4) \rangle\\
	F_{\phi\eta}(\tau_1,\tau_2,\tau_3,\tau_4)\equiv & {1\over NM} \langle \phi_a(\tau_1)\phi_a(\tau_2)\chi_{k,\sigma_3}(\tau_3)\chi_{k,\sigma_4}(\tau_4) \rangle\\
	F_{\phi\phi}(\tau_1,\tau_2,\tau_3,\tau_4)\equiv & {1\over M^2} \langle \phi_a(\tau_1)\phi_a(\tau_2)\phi_b(\tau_3)\phi_b(\tau_4) \rangle
\end{align}
For this, we expand the bi-local fields $\Psi$ and $\Phi$ around the classical solution in large $N,M$:
\begin{align}
	\Psi(\tau_1,\sigma_1;\tau_2,\sigma_2)=&\Psi_0(\tau_1,\sigma_1;\tau_2,\sigma_2)+ {1\over \sqrt{N}} \eta_{\sigma_1\sigma_2}(\tau_1,\tau_2)\\
	\Phi(\tau_1,\tau_2)=&\Phi_0(\tau_1,\tau_2) + {1\over \sqrt{M} } \phi(\tau_1,\tau_2) 
\end{align}
Accordingly, from the large-$N$ expansion of the collective action in Eq.~\eqref{eq: collective action q}
\begin{align}
	S_{col}=N S_{col}^{(0)}+ S_{col}^{(2)}+ {1\over \sqrt{N}} S_{col}^{(3)}+\cdots
\end{align}
we can read the quadratic action
%
%
%
\begin{align}
	S_{col}^{(2)}\,=\,& -{1\over 4} \Tr \left( \Psi_0^{-1} \star \eta \star \Psi_0^{-1}\star  \eta  \right) + {1\over 4} \tr \left( \Phi_0^{-1}\star \phi \star \Phi_0^{-1}\star \phi \right)\cr
	&+ {(q-1)J^2\over 4} \sum_{X_1,X_2}  \Phi_0(\tau_1,\tau_2)[\Psi_0(X_1,X_2)]^{q-2} \eta(X_1,X_2)\eta(X_1,X_2)\cr
	&+ {J^2\over 2\sqrt{r}} \int d\tau_1d\tau_2\; [\Psi_0(X_1,X_2)]^{q-1} \phi(\tau_1,\tau_2)\eta(X_1,X_2)\ ,
\end{align}
where $X$ denotes a collective coordinate of $(\tau, \sigma)$ and the summation over $X$ denotes the integration over $\tau$ together with summation over $\sigma$. For the Schwinger-Dyson equation for the four point functions, we will use the following path integral identity.
\begin{align}
	0\,=\,&\int \cD\eta \cD \phi \; {\delta \over \delta \eta(X_6,X_5) } \left[ \eta(X_3,X_4) e^{-S_{col}^{(2)}}\right]\\
	=\,&{1\over 2}\left[ \delta(X_{36})\delta(X_{45})-\delta(X_{35})\delta(X_{46}) \right]+{1\over 2} \sum_{X_1,X_2} \Psi_0^{-1}(X_5,X_1)\Psi_0^{-1}(X_2,X_6)F_{\eta\eta}(X_1,X_2,X_3,X_4)\cr
	&+ {(q-1)J^2 \over 2 }\Phi_0(\tau_5,\tau_6)[\Psi_0(X_5,X_6)]^{q-2}F_{\eta\eta}(\tau_5,\tau_6,\tau_3,\tau_4)+ {J^2\over 2\sqrt{r}} [\Psi_0(X_5,X_6)]^{q-1}F_{\phi \eta}(\tau_5,\tau_6,X_3,X_4)\ .
\end{align}
By multiplying $\displaystyle \sum_{X_5,X_6} \Psi(X_1,X_5)\Psi(X_6,X_2)$, we have the Schwinger-Dyson equation for the four point function.
\begin{align}
	&F_{\eta\eta}(X_1,X_2,X_3,X_4)- {J^2\over \sqrt{r}} \sum_{X_5,X_6}  \Psi_0(X_1,X_5)\Psi_0(X_2,X_6) [\Psi_0(X_5,X_6)]^{q-1}F_{\phi \eta}(\tau_5,\tau_6,X_3,X_4)\cr
	&-  (q-1)J^2 \sum_{X_5,X_6}  \Psi_0(X_1,X_5)\Psi_0(X_2,X_6) \Phi_0(\tau_5,\tau_6)[\Psi_0(X_5,X_6)]^{q-2}F_{\eta\eta}(X_5,X_6,X_3,X_4)\cr
	=\,&\Psi_0(X_1,X_3)\Psi_0(X_4,X_2)-\Psi_0(X_1,X_4)\Psi_0(X_3,X_2)\ .
\end{align}
In the same way, one can generate other Schwinger-Dyson equations.
%
%
%
\begin{align}
	&F_{\eta\phi}(X_1,X_2,\tau_3,\tau_4)- {J^2\over \sqrt{r}} \sum_{X_5,X_6}   \Psi_0(X_1,X_5)\Psi_0(X_2,X_6) [\Psi_0(X_5,X_6)]^{q-1}F_{\phi \phi}(\tau_5,\tau_6,\tau_3,\tau_4)\cr
	&- (q-1)J^2\sum_{X_5,X_6}   \Psi_0(X_1,X_5)\Psi_0(X_2,X_6)\Phi_0(\tau_5,\tau_6)[\Psi_0(X_5,X_6)]^{q-2}F_{\eta\phi}(X_5,X_6,\tau_3,\tau_4)\,=\,0\ ,
\end{align}
%
%
%
%
\begin{align}
	F_{\phi \eta }(\tau_1,\tau_2,X_3,X_4)\,=\,- {J^2\over \sqrt{r}} \sum_{X_5,X_6}  \Phi_0(\tau_1,\tau_5)\Phi_0(\tau_2,\tau_6)[\Psi_0(X_5,X_6)]^{q-1}F_{\eta \eta}(X_5,X_6,X_3,X_4)\ ,
\end{align}
%
%
%
%
%
\begin{align}
	F_{\phi \phi }(\tau_1,\tau_2,\tau_3,\tau_4)\,=\,&\Phi_0(\tau_1,\tau_4)\Phi_0(\tau_2,\tau_3)+\Phi_0(\tau_1,\tau_3)\Phi_0(\tau_2,\tau_4)\cr
	&- {J^2\over \sqrt{r}} \sum_{X_5,X_6}  \Phi_0(\tau_1,\tau_5)\Phi_0(\tau_2,\tau_6)[\Psi_0(X_5,X_6)]^{q-1}F_{\eta \phi}(X_5,X_6,\tau_3,\tau_4)\ .
\end{align}
Those four Schwinger-Dyson equations can also be written in the following compact form.
\begin{align}
	\begin{pmatrix}
	F_{\eta\eta} & F_{\eta \phi}\\
	F_{\phi\eta} & F_{\phi \phi}\\
	\end{pmatrix}\, = \, 	\begin{pmatrix}
	F_{0,\eta\eta} & 0\\
	0 & F_{0,\phi \phi}\\
	\end{pmatrix} +\begin{pmatrix}
	K_{\eta\eta} & K_{\eta \phi}\\
	K_{\phi \eta} & 0 \\
	\end{pmatrix} \star	\begin{pmatrix}
	F_{\eta\eta} & F_{\eta \phi}\\
	F_{\phi\eta} & F_{\phi \phi}\\
	\end{pmatrix}\ ,
\end{align}
%
%
This can be a matrix geometric series where the common ratio $K$'s and the initial term $F_0$'s are defined by
\begin{align}
	K_{\eta \eta} (X_1,X_2,X_5,X_6)\,=\,&  (q-1)J^2    \Psi_0(X_1,X_5)\Psi_0(X_2,X_6) \Phi_0(\tau_5,\tau_6)[\Psi_0(X_5,X_6)]^{q-2}\ ,\\
	K_{\eta \phi} (X_1,X_2,X_5,X_6)\,=\,& {J^2\over \sqrt{r}}    \Psi_0(X_1,X_5)\Psi_0(X_2,X_6) [\Psi_0(X_5,X_6)]^{q-1} \ ,\\
	K_{\phi \eta} (X_1,X_2,X_5,X_6)\,=\,&- {J^2\over \sqrt{r}}  \Phi_0(\tau_1,\tau_5)\Phi_0(\tau_2,\tau_6)[\Psi_0(X_5,X_6)]^{q-1}\ ,\\
	K_{\phi \phi} (X_1,X_2,X_5,X_6)\,=\,&0 \ ,
\end{align}
and
\begin{align}
	F_{0,\eta\eta}(X_1,X_2,X_3,X_4)\,=\,&\Psi_0(X_1,X_3)\Psi_0(X_4,X_2)-\Psi_0(X_1,X_4)\Psi_0(X_3,X_2)\ ,\\
	F_{0,\phi\phi}(\tau_1,\tau_2,\tau_3,\tau_4)\,=\,&\Phi_0(\tau_1,\tau_4)\Phi_0(\tau_2,\tau_3)+\Phi_0(\tau_1,\tau_3)\Phi_0(\tau_2,\tau_4)\ .
\end{align}
Using the classical solution of the bi-locals, the kernels $K$'s can be written as follows.
\begin{align}
	K_{\eta \eta} (X_1,X_2,X_3,X_4)\,=\,&  (q-1)J^2 b_{\Psi}^q b_\Phi { \sgn(\tau_{13}) \sgn(\tau_{24})  \over |\tau_{13}\tau_{24}\tau_{34}^{q-2}|^{2\Delta_\Psi}|\tau_{34}|^{2\Delta_\Phi}} \qquad\mbox{for} \;\;(\sigma_1,\sigma_2,\sigma_3,\sigma_4)\,=\,(\pm,\pm,\pm,\pm)\ ,\\
	K_{\eta \phi} (X_1,X_2,X_3,X_4)\,=\,& {J^2\over \sqrt{r}}  b_{\Psi}^{q+1}   { \sgn(\tau_{13}) \sgn(\tau_{24})\sgn(\tau_{34})  \over |\tau_{13}\tau_{24}\tau_{34}^{q-1}|^{2\Delta_\Psi} }\qquad\mbox{for} \;\; (\sigma_1,\sigma_2
	)\,=\,(\pm,\pm) \ ,\\
	K_{\phi \eta} (X_1,X_2,X_3,X_4)\,=\,&- {J^2\over \sqrt{r}}  b_{\Psi}^{q-1} b_\Phi^2    {  \sgn(\tau_{34})  \over |\tau_{34}^{q-1}|^{2\Delta_\Psi}|\tau_{13}\tau_{24} |^{2\Delta_\Phi}} \qquad\mbox{for} \;\; (\sigma_3,\sigma_4)\,=\,(\pm,\pm)  \ ,\\
	K_{\phi \phi} (X_1,X_2,X_3,X_4)\,=\,&0\ .
\end{align}
Now, we diagonalize the kernels $K$'s by the conformal partial wave function to evaluate the four point function. For this, we consider operators $\cO_h$ of conformal dimension $h$ which appears in the decomposition of bi-local operators. Then, we define $\Upsilon^{\eta,\phi}_h(\tau_1,\tau_2)$ by the overlap between the bi-local operator and the operator $\cO_h$.
\begin{align}
    \Upsilon^\eta_h(\tau_1,\tau_2) \,\sim\, \langle \Psi(\tau_1,+;\tau_2,+)\cO_h \rangle + \langle \Psi(\tau_1,-;\tau_2,-)\cO_h \rangle\;\;,\quad \Upsilon^\phi_h(\tau_1,\tau_2)\,\sim\, \langle \Phi(\tau_1,\tau_2)\cO_h \rangle\ . 
\end{align}
%
Using the universal form of the three point function, they can be written as follows with suitable normalization.
\begin{align}
	\Upsilon^\eta_h(\tau_1,
	\tau_2)\,=\,{\sgn(\tau_{12})\over |\tau_{12}|^{2\Delta_\Psi -h}}\;\;, \quad \Upsilon^\phi_h(\tau_1,
	\tau_2)\,=\,{1\over |\tau_{12}|^{2\Delta_\Phi -h}} \ .
\end{align}
Note that $\Upsilon^\eta_h$ and $\Upsilon^\phi_h$ diagonalize the conformal Casimir operator by construction, and they also simultaneously diagonalize the kernels $K$'s because the kernel and conformal Casimir operator commute. 
\begin{align}
	(K_{\eta\eta}\star \Upsilon^\eta_h)(X_1,X_2)\,=\,& k_{\eta\eta} \Upsilon^\eta_h(X_1,X_2)\ ,\\
	(K_{\phi\eta}\star \Upsilon^\eta_h)(X_1,X_2)\,=\,&k_{\phi \eta}  \Upsilon^{\phi}_h(X_1,X_2)\ , \\
	(K_{\eta\phi}\star \Upsilon^{\phi}_h)(X_1,X_2)\,=\,& k_{ \eta \phi } \Upsilon^\eta_h(X_1,X_2)\ .
\end{align}
To evaluate the eigenvalue $k_{\eta\eta}\,,\; k_{\phi\eta}$ and $k_{\eta\phi}$, we use the following integral identities
\begin{align}
	&(q-1)J^2b_{\Psi}^q b_\Phi \int_{-\infty}^\infty d\tau_3d \tau_4 \;  { \sgn(\tau_{13}) \sgn(\tau_{24})\sgn(\tau_{34})  \over |\tau_{13}|^{2\Delta_\Psi}  |\tau_{24}|^{2\Delta_\Psi} |\tau_{34}|^{2-2\Delta_\Psi-h} }\cr 
	=\,&(q-1)J^2b_{\Psi}^q b_\Phi  {\pi^2 \Gamma(2\Delta_\Psi-h) \sin {\pi(2\Delta_\Psi -h) \over 2} \over \sin^2{\pi \Delta_\Psi}\sin{ \pi(2\Delta_\Psi+h) \over 2} \Gamma(2-2\Delta_\Psi-h)[\Gamma(2\Delta_\Psi)]^2} {\sgn(\tau_{12})\over |\tau_{12}|^{2\Delta_\Psi-h}}\ ,\\
	&- 2 {J^2\over \sqrt{r}}  b_{\Psi}^{q-1} b_\Phi^2   \int_{-\infty}^\infty d\tau_3 d\tau_4 \; {1  \over |\tau_{13}|^{2\Delta_\Phi} |\tau_{24} |^{2\Delta_\Phi} |\tau_{34}|^{2-2\Delta_\Phi-h}} \cr
	=\,&  2 {J^2\over \sqrt{r}}  b_{\Psi}^{q-1} b_\Phi^2  {\pi^2 \Gamma(2\Delta_\Phi-h) \cos {\pi(2\Delta_\Phi -h) \over 2} \over \cos^2{\pi \Delta_\Phi}\cos{ \pi(2\Delta_\Phi+h) \over 2} \Gamma(2-2\Delta_\Phi-h)[\Gamma(2\Delta_\Phi)]^2} {\sgn(\tau_{12})\over |\tau_{12}|^{2\Delta_\Phi-h}}\ ,\\
	 &{J^2\over \sqrt{r}}   b_{\Psi}^{q+1} \int d\tau_3d\tau_4\; { \sgn(\tau_{13}) \sgn(\tau_{24})\sgn(\tau_{34})  \over |\tau_{13}|^{2\Delta_\Psi} |\tau_{24}|^{2\Delta_\Psi} |\tau_{34}|^{2-2\Delta_\Psi-h} } \cr
	 =\,& {J^2\over \sqrt{r}}   b_{\Psi}^{q+1}  {\pi^2 \Gamma(2\Delta_\Psi-h) \sin {\pi(2\Delta_\Psi -h) \over 2} \over \sin^2{\pi \Delta_\Psi}\sin{ \pi(2\Delta_\Psi+h) \over 2} \Gamma(2-2\Delta_\Psi-h)[\Gamma(2\Delta_\Psi)]^2} {\sgn(\tau_{12})\over |\tau_{12}|^{2\Delta_\Psi-h}}\ .
\end{align}
Then, we obtain 
\begin{align}
	k_{\eta\eta}\,=\,&(q-1)J^2b_{\Psi}^q b_\Phi  {\pi^2 \Gamma(2\Delta_\Psi-h) \sin {\pi(2\Delta_\Psi -h) \over 2} \over \sin^2{\pi \Delta_\Psi}\sin{ \pi(2\Delta_\Psi+h) \over 2} \Gamma(2-2\Delta_\Psi-h)[\Gamma(2\Delta_\Psi)]^2}\ ,\\
	k_{\phi\eta}\,=\,& 2 {J^2\over \sqrt{r}}  b_{\Psi}^{q-1} b_\Phi^2  {\pi^2 \Gamma(2\Delta_\Phi-h) \cos {\pi(2\Delta_\Phi -h) \over 2} \over \cos^2{\pi \Delta_\Phi}\cos{ \pi(2\Delta_\Phi+h) \over 2} \Gamma(2-2\Delta_\Phi-h)[\Gamma(2\Delta_\Phi)]^2}\ , \\
	k_{\eta\phi}\,=\,&	 {J^2\over \sqrt{r}}  b_{\Psi}^{q+1}  {\pi^2 \Gamma(2\Delta_\Psi-h) \sin {\pi(2\Delta_\Psi -h) \over 2} \over \sin^2{\pi \Delta_\Psi}\sin{ \pi(2\Delta_\Psi+h) \over 2} \Gamma(2-2\Delta_\Psi-h)[\Gamma(2\Delta_\Psi)]^2} \ .
\end{align}
In the basis $\Upsilon^\eta_h$ and $\Upsilon^\phi_h$, the common ratio of the geometric series for the four point function can be represented by
\begin{align}
	\begin{pmatrix}
	k_{\eta\eta} & k_{\eta\phi}\\
	k_{\phi \eta} & 0\\
	\end{pmatrix}
\end{align}
and the conformal dimensions of the operators which appears in the OPE limit of four point function can be found by the following equation:
\begin{align}
	\det\left[\begin{pmatrix}
	k_{\eta\eta} & k_{\eta\phi}\\
	k_{\phi \eta} & 0\\
	\end{pmatrix}- \begin{pmatrix}
	1 & 0 \\
	0 & 1\\
	\end{pmatrix}\right]=0
\end{align}
For $q=2$ and $r=1$, the dimensions of the operators in the intermediate channel of four point functions are
\begin{align}
    h=1\;,\;\; 2\;,\;\; 3.05788\cdots\;,\;\; 3.85483\cdots\;,\;\; 5.08416\;,\;\; \cdots
\end{align}
Note $h=1\;,\;2$ mode will lead to divergence in the strict conformal limit.

\subsection{Out-of-time-ordered Correlator}
\label{sec: otoc}

In this appendix, we will evaluate the Lyapunov exponent of the out-of-time-ordered correlator~(OTOC)
\begin{align}
    F_{\eta\eta}(t_1,t_2)\,\equiv\,&{1\over N^2} \sum_{i,j=1}^N \sum_{\sigma,\rho=\pm} \tr\bigg[e^{-{\beta H\over 4}} \chi_{j,\sigma}(t_1) e^{-{\beta H\over 4}}  \chi_{k,\rho }(0) e^{-{\beta H\over 4}} \chi_{j,\sigma}(t_2) e^{-{\beta H\over 4}}  \chi_{k,\rho}(0) e^{-{\beta H\over 4}}\bigg]\ ,\\
    F_{\phi\eta}(t_1,t_2)\,\equiv\,&{1\over MN} \sum_{i=1}^N \sum_{\sigma=\pm }\sum_{a=1}^M\tr\bigg[e^{-{\beta H\over 4}} \phi_a(t_1) e^{-{\beta H\over 4}}  \chi_{j,\sigma }(0) e^{-{\beta H\over 4}} \phi_a(t_2) e^{-{\beta H\over 4}}  \chi_{j,\sigma}(0) e^{-{\beta H\over 4}}\bigg]\ ,\\
    F_{\phi\phi}(t_1,t_2)\,\equiv\,&{1\over M^2}\sum_{a,b=1}^M\tr\bigg[e^{-{\beta H\over 4}} \phi_a(t_1) e^{-{\beta H\over 4}} \phi_b(0) e^{-{\beta H\over 4}} \phi_a(t_2) e^{-{\beta H\over 4}}  \phi_b(0) e^{-{\beta H\over 4}}\bigg]\ ,
\end{align}
by using the retarded kernel in the real-time formulation~\cite{Maldacena:2016hyu}. For this, we calculate the retarded two point function and Wightman function by a proper Wick rotation.
\begin{align}
	\Psi_R(X_1,X_2)\,=\,& 2\cos (\pi \Delta_\Psi) b_{\sigma_1\sigma_2}  \left({ \pi \over \beta \sinh {\pi t_{12}\over \beta} }\right)^{2\Delta_\Psi}\theta(t_{12})\ ,\\
	\Phi_R(t_1,t_2)\,=\,& -2i\sin (\pi \Delta_\Phi) b_\Phi  \left({ \pi \over \beta \sinh {\pi t_{12}\over \beta} }\right)^{2\Delta_\Phi}\theta(t_{12})\ ,\\
	\Psi_{lr}(X_1,X_2)=& b_{\sigma_1\sigma_2} \left({ \pi \over \beta \cosh {\pi t_{12}\over \beta} }\right)^{2\Delta_\Psi} \ , \\
	\Phi_{lr}(X_1,X_2)=&b_\Phi  \left({ \pi \over \beta \cosh {\pi t_{12}\over \beta} }\right)^{2\Delta_\Phi}\ .
\end{align}
Using them, we can obtain the retarded kernels~\cite{Maldacena:2016hyu}.
\begin{align}
	&K^R_{\eta \eta} (X_1,X_2,X_3,X_4)\,=\,-  (q-1)J^2 b_{\Psi}^q b_\Phi  4\cos^2 (\pi \Delta_\Psi) \cr
	&\times \left({ \pi \over \beta \sinh {\pi t_{13}\over \beta} }\right)^{2\Delta_\Psi} \left({ \pi \over \beta \sinh {\pi t_{24}\over \beta} }\right)^{2\Delta_\Psi}\left({ \pi \over \beta \cosh {\pi t_{34}\over \beta} }\right)^{2\Delta_\Phi}  \left({ \pi \over \beta \cosh {\pi t_{34}\over \beta} }\right)^{2\Delta_\Psi(q-2)}   \theta(t_{13}) \theta(t_{24})\cr
	& \qquad\mbox{for} \;\; (\pm,\pm,\pm,\pm)\ , \\
	&K^R_{\eta \phi} (X_1,X_2,X_3,X_4)\,=\,- {J^2\over \sqrt{r}}  b_{\Psi}^{q+1}   4\cos^2 (\pi \Delta_\Psi) \cr
	&\times \left({ \pi \over \beta \sinh {\pi t_{13}\over \beta} }\right)^{2\Delta_\Psi} \left({ \pi \over \beta \sinh {\pi t_{24}\over \beta} }\right)^{2\Delta_\Psi}  \left({ \pi \over \beta \cosh {\pi t_{34}\over \beta} }\right)^{2\Delta_\Psi(q-1)}   \theta(t_{13}) \theta(t_{24})\ ,\\
	&K^R_{\phi \eta} (X_1,X_2,X_3,X_4)\,=\,- {J^2\over \sqrt{r}}  b_{\Psi}^{q-1} b_\Phi^2 4\sin^2 (\pi \Delta_\Phi) \cr
	& \times \left({ \pi \over \beta \sinh {\pi t_{13}\over \beta} }\right)^{2\Delta_\Phi} \left({ \pi \over \beta \sinh {\pi t_{24}\over \beta} }\right)^{2\Delta_\Phi}  \left({ \pi \over \beta \cosh {\pi t_{34}\over \beta} }\right)^{2\Delta_\Psi(q-1)}   \theta(t_{13}) \theta(t_{24}) \cr
	&    \qquad\mbox{for} \;\; (\pm,\pm)\ ,  \\
	&K^R_{\phi \phi} (X_1,X_2,X_3,X_4)\,=\,0 \ .
\end{align}
And the Schwinger-Dyson equation for the OTOC is given by
\begin{align}
	\begin{pmatrix}
	F_{\eta\eta} & F_{\eta \phi}\\
	F_{\phi\eta} & F_{\phi \phi}\\
	\end{pmatrix}\, = \, \begin{pmatrix}
	K^R_{\eta\eta} & K^R_{\eta \phi}\\
	K^R_{\phi \eta} & 0 \\
	\end{pmatrix} \star	\begin{pmatrix}
	F_{\eta\eta} & F_{\eta \phi}\\
	F_{\phi\eta} & F_{\phi \phi}\\
	\end{pmatrix}\ ,
\end{align}
We take the following ansatz for the OTOCs
\begin{align}
	\begin{pmatrix}
	F_{\eta\eta} & F_{\eta \phi}\\
	F_{\phi\eta} & F_{\phi \phi}\\
	\end{pmatrix}\, \sim\, \Upsilon_h(t_1,t_2)\,\equiv \, {e^{-{\pi h\over \beta}(t_1+t_2)}\over \left[\cosh{\pi t_{12}\over \beta}\right]^{2\Delta_\psi-h}}\ ,
\end{align}
where the function $\Upsilon_h(t_1,t_2)$ diagonalizes the retarded kernels. 
\begin{align}
	(K^R_{\eta\eta}\star \Upsilon_h)(t_1,t_2)\,=\,& k^R_{\eta\eta} \Upsilon_h(t_1,t_2)\ ,\\
	(K^R_{\phi\eta}\star \Upsilon_h)(t_1,t_2)\,=\,&k^R_{\phi \eta}  \Upsilon_h(t_1,t_2)\ , \\
	(K^R_{\eta\phi}\star \Upsilon_h)(t_1,t_2)\,=\,& k^R_{ \eta \phi } \Upsilon_h(t_1,t_2)\ .
\end{align}
Using the conformal dimension $\Delta_\Psi$ from Eq.~\eqref{eq: conformal dim eq q 2}, we can obtain the eigenvalue $k^R$'s.
\begin{align}
	k^R_{\eta\eta}\,=\,&-(q-1)J^2 b_{\Psi}^q b_\Phi  4\cos^2 (\pi \Delta_\Psi){\left[\Gamma(1-2\Delta_\Psi)\right]^2\Gamma(2\Delta_\Psi-h)\over \Gamma (2-h - 2\Delta_\Psi)} \ ,\\
	k^R_{\phi\eta}\,=\,&-2 {J^2\over \sqrt{r}}  b_{\Psi}^{q-1} b_\Phi^2 4\sin^2 (\pi \Delta_\Phi)  \left({\pi \over \beta}\right)^{2\Delta_\Phi-2\Delta_\Psi}    {\left[\Gamma(1-2\Delta_\Phi)\right]^2\Gamma(2\Delta_\Phi-h)\over \Gamma (2-h -2\Delta_\Phi)} \ ,\\
	k^R_{\eta\phi}\,=\,&-{J^2\over \sqrt{r}}  b_{\Psi}^{q+1}   4\cos^2 (\pi \Delta_\Psi) \left({\pi \over \beta}\right)^{-2\Delta_\Phi+2\Delta_\Psi}    {\left[\Gamma(1-2\Delta_\Psi)\right]^2\Gamma(2\Delta_\Psi-h)\over \Gamma (2-h-2\Delta_\Psi)} \ , \\
	k^R_{\phi\phi}\,=\,&0\ .
\end{align}
For a solution of the Schwinger-Dyson equation for the OTOC with retarded kernel, the common ratio of the matrix geometric series should have the eigenvalue 1~\cite{Maldacena:2016hyu}, and we can confirm that 
\begin{align}
	\det\left[\begin{pmatrix}
	k^R_{\eta\eta} & k^R_{\eta^S\phi}\\
	k^R_{\phi \eta^S} & 0\\
	\end{pmatrix}- \begin{pmatrix}
	1 & 0 \\
	0 & 1\\
	\end{pmatrix}\right]=0\hspace{8mm}\mbox{for}\;\; h=-1 .
\end{align}
This implies that the OTOC $\Upsilon_h(t_1,t_2)$ grows exponentially in time
\begin{align}
    \Upsilon_{h=-1} (t_1,t_2)\sim e^{{2\pi\over \beta} t}\hspace{5mm}\text{where}\quad t\equiv {1\over 2}(t_1+t_2)\ ,
\end{align}
and one can read off the Lyapunov exponent $\lambda_L={2\pi \over \beta}$. This proves that our model saturates the chaos bound.

\end{widetext}

\bibliography{QMC_SYK}

\begin{thebibliography}{60}%
\makeatletter
\providecommand \@ifxundefined [1]{%
 \@ifx{#1\undefined}
}%
\providecommand \@ifnum [1]{%
 \ifnum #1\expandafter \@firstoftwo
 \else \expandafter \@secondoftwo
 \fi
}%
\providecommand \@ifx [1]{%
 \ifx #1\expandafter \@firstoftwo
 \else \expandafter \@secondoftwo
 \fi
}%
\providecommand \natexlab [1]{#1}%
\providecommand \enquote  [1]{``#1''}%
\providecommand \bibnamefont  [1]{#1}%
\providecommand \bibfnamefont [1]{#1}%
\providecommand \citenamefont [1]{#1}%
\providecommand \href@noop [0]{\@secondoftwo}%
\providecommand \href [0]{\begingroup \@sanitize@url \@href}%
\providecommand \@href[1]{\@@startlink{#1}\@@href}%
\providecommand \@@href[1]{\endgroup#1\@@endlink}%
\providecommand \@sanitize@url [0]{\catcode `\\12\catcode `\$12\catcode
  `\&12\catcode `\#12\catcode `\^12\catcode `\_12\catcode `\%12\relax}%
\providecommand \@@startlink[1]{}%
\providecommand \@@endlink[0]{}%
\providecommand \url  [0]{\begingroup\@sanitize@url \@url }%
\providecommand \@url [1]{\endgroup\@href {#1}{\urlprefix }}%
\providecommand \urlprefix  [0]{URL }%
\providecommand \Eprint [0]{\href }%
\providecommand \doibase [0]{https://doi.org/}%
\providecommand \selectlanguage [0]{\@gobble}%
\providecommand \bibinfo  [0]{\@secondoftwo}%
\providecommand \bibfield  [0]{\@secondoftwo}%
\providecommand \translation [1]{[#1]}%
\providecommand \BibitemOpen [0]{}%
\providecommand \bibitemStop [0]{}%
\providecommand \bibitemNoStop [0]{.\EOS\space}%
\providecommand \EOS [0]{\spacefactor3000\relax}%
\providecommand \BibitemShut  [1]{\csname bibitem#1\endcsname}%
\let\auto@bib@innerbib\@empty
\bibitem [{\citenamefont {Stewart}(2001)}]{RevModPhys.73.797}%
  \BibitemOpen
  \bibfield  {author} {\bibinfo {author} {\bibfnamefont {G.~R.}\ \bibnamefont
  {Stewart}},\ }\bibfield  {title} {\bibinfo {title} {{Non-Fermi-liquid
  behavior in $d$- and $f$-electron metals}},\ }\href
  {https://doi.org/10.1103/RevModPhys.73.797} {\bibfield  {journal} {\bibinfo
  {journal} {Rev. Mod. Phys.}\ }\textbf {\bibinfo {volume} {73}},\ \bibinfo
  {pages} {797} (\bibinfo {year} {2001})}\BibitemShut {NoStop}%
\bibitem [{\citenamefont {Troyer}\ and\ \citenamefont
  {Wiese}(2005)}]{PhysRevLett.94.170201}%
  \BibitemOpen
  \bibfield  {author} {\bibinfo {author} {\bibfnamefont {M.}~\bibnamefont
  {Troyer}}\ and\ \bibinfo {author} {\bibfnamefont {U.-J.}\ \bibnamefont
  {Wiese}},\ }\bibfield  {title} {\bibinfo {title} {{Computational Complexity
  and Fundamental Limitations to Fermionic Quantum Monte Carlo Simulations}},\
  }\href {https://doi.org/10.1103/PhysRevLett.94.170201} {\bibfield  {journal}
  {\bibinfo  {journal} {Phys. Rev. Lett.}\ }\textbf {\bibinfo {volume} {94}},\
  \bibinfo {pages} {170201} (\bibinfo {year} {2005})}\BibitemShut {NoStop}%
\bibitem [{\citenamefont {Scalapino}\ and\ \citenamefont
  {Sugar}(1981{\natexlab{a}})}]{PhysRevLett.46.519}%
  \BibitemOpen
  \bibfield  {author} {\bibinfo {author} {\bibfnamefont {D.~J.}\ \bibnamefont
  {Scalapino}}\ and\ \bibinfo {author} {\bibfnamefont {R.~L.}\ \bibnamefont
  {Sugar}},\ }\bibfield  {title} {\bibinfo {title} {{Method for Performing
  Monte Carlo Calculations for Systems with Fermions}},\ }\href
  {https://doi.org/10.1103/PhysRevLett.46.519} {\bibfield  {journal} {\bibinfo
  {journal} {Phys. Rev. Lett.}\ }\textbf {\bibinfo {volume} {46}},\ \bibinfo
  {pages} {519} (\bibinfo {year} {1981}{\natexlab{a}})}\BibitemShut {NoStop}%
\bibitem [{\citenamefont {Blankenbecler}\ \emph {et~al.}(1981)\citenamefont
  {Blankenbecler}, \citenamefont {Scalapino},\ and\ \citenamefont
  {Sugar}}]{PhysRevD.24.2278}%
  \BibitemOpen
  \bibfield  {author} {\bibinfo {author} {\bibfnamefont {R.}~\bibnamefont
  {Blankenbecler}}, \bibinfo {author} {\bibfnamefont {D.~J.}\ \bibnamefont
  {Scalapino}},\ and\ \bibinfo {author} {\bibfnamefont {R.~L.}\ \bibnamefont
  {Sugar}},\ }\bibfield  {title} {\bibinfo {title} {{Monte Carlo calculations
  of coupled boson-fermion systems. I}},\ }\href
  {https://doi.org/10.1103/PhysRevD.24.2278} {\bibfield  {journal} {\bibinfo
  {journal} {Phys. Rev. D}\ }\textbf {\bibinfo {volume} {24}},\ \bibinfo
  {pages} {2278} (\bibinfo {year} {1981})}\BibitemShut {NoStop}%
\bibitem [{\citenamefont {Scalapino}\ and\ \citenamefont
  {Sugar}(1981{\natexlab{b}})}]{PhysRevB.24.4295}%
  \BibitemOpen
  \bibfield  {author} {\bibinfo {author} {\bibfnamefont {D.~J.}\ \bibnamefont
  {Scalapino}}\ and\ \bibinfo {author} {\bibfnamefont {R.~L.}\ \bibnamefont
  {Sugar}},\ }\bibfield  {title} {\bibinfo {title} {{Monte Carlo calculations
  of coupled boson-fermion systems. II}},\ }\href
  {https://doi.org/10.1103/PhysRevB.24.4295} {\bibfield  {journal} {\bibinfo
  {journal} {Phys. Rev. B}\ }\textbf {\bibinfo {volume} {24}},\ \bibinfo
  {pages} {4295} (\bibinfo {year} {1981}{\natexlab{b}})}\BibitemShut {NoStop}%
\bibitem [{\citenamefont {Hirsch}\ \emph {et~al.}(1981)\citenamefont {Hirsch},
  \citenamefont {Scalapino}, \citenamefont {Sugar},\ and\ \citenamefont
  {Blankenbecler}}]{PhysRevLett.47.1628}%
  \BibitemOpen
  \bibfield  {author} {\bibinfo {author} {\bibfnamefont {J.~E.}\ \bibnamefont
  {Hirsch}}, \bibinfo {author} {\bibfnamefont {D.~J.}\ \bibnamefont
  {Scalapino}}, \bibinfo {author} {\bibfnamefont {R.~L.}\ \bibnamefont
  {Sugar}},\ and\ \bibinfo {author} {\bibfnamefont {R.}~\bibnamefont
  {Blankenbecler}},\ }\bibfield  {title} {\bibinfo {title} {{Efficient Monte
  Carlo Procedure for Systems with Fermions}},\ }\href
  {https://doi.org/10.1103/PhysRevLett.47.1628} {\bibfield  {journal} {\bibinfo
   {journal} {Phys. Rev. Lett.}\ }\textbf {\bibinfo {volume} {47}},\ \bibinfo
  {pages} {1628} (\bibinfo {year} {1981})}\BibitemShut {NoStop}%
\bibitem [{\citenamefont {Hirsch}\ \emph {et~al.}(1982)\citenamefont {Hirsch},
  \citenamefont {Sugar}, \citenamefont {Scalapino},\ and\ \citenamefont
  {Blankenbecler}}]{PhysRevB.26.5033}%
  \BibitemOpen
  \bibfield  {author} {\bibinfo {author} {\bibfnamefont {J.~E.}\ \bibnamefont
  {Hirsch}}, \bibinfo {author} {\bibfnamefont {R.~L.}\ \bibnamefont {Sugar}},
  \bibinfo {author} {\bibfnamefont {D.~J.}\ \bibnamefont {Scalapino}},\ and\
  \bibinfo {author} {\bibfnamefont {R.}~\bibnamefont {Blankenbecler}},\
  }\bibfield  {title} {\bibinfo {title} {{Monte Carlo simulations of
  one-dimensional fermion systems}},\ }\href
  {https://doi.org/10.1103/PhysRevB.26.5033} {\bibfield  {journal} {\bibinfo
  {journal} {Phys. Rev. B}\ }\textbf {\bibinfo {volume} {26}},\ \bibinfo
  {pages} {5033} (\bibinfo {year} {1982})}\BibitemShut {NoStop}%
\bibitem [{\citenamefont {Hirsch}(1983)}]{PhysRevB.28.4059}%
  \BibitemOpen
  \bibfield  {author} {\bibinfo {author} {\bibfnamefont {J.~E.}\ \bibnamefont
  {Hirsch}},\ }\bibfield  {title} {\bibinfo {title} {{Discrete
  Hubbard-Stratonovich transformation for fermion lattice models}},\ }\href
  {https://doi.org/10.1103/PhysRevB.28.4059} {\bibfield  {journal} {\bibinfo
  {journal} {Phys. Rev. B}\ }\textbf {\bibinfo {volume} {28}},\ \bibinfo
  {pages} {4059} (\bibinfo {year} {1983})}\BibitemShut {NoStop}%
\bibitem [{\citenamefont {Wu}\ and\ \citenamefont
  {Zhang}(2005)}]{PhysRevB.71.155115}%
  \BibitemOpen
  \bibfield  {author} {\bibinfo {author} {\bibfnamefont {C.}~\bibnamefont
  {Wu}}\ and\ \bibinfo {author} {\bibfnamefont {S.-C.}\ \bibnamefont {Zhang}},\
  }\bibfield  {title} {\bibinfo {title} {{Sufficient condition for absence of
  the sign problem in the fermionic quantum Monte Carlo algorithm}},\ }\href
  {https://doi.org/10.1103/PhysRevB.71.155115} {\bibfield  {journal} {\bibinfo
  {journal} {Phys. Rev. B}\ }\textbf {\bibinfo {volume} {71}},\ \bibinfo
  {pages} {155115} (\bibinfo {year} {2005})}\BibitemShut {NoStop}%
\bibitem [{\citenamefont {Sandvik}(2010)}]{sandvik2010computational}%
  \BibitemOpen
  \bibfield  {author} {\bibinfo {author} {\bibfnamefont {A.~W.}\ \bibnamefont
  {Sandvik}},\ }\bibfield  {title} {\bibinfo {title} {{Computational studies of
  quantum spin systems}},\ }in\ \href {https://doi.org/10.1063/1.3518900}
  {\emph {\bibinfo {booktitle} {AIP Conference Proceedings}}},\ Vol.\ \bibinfo
  {volume} {1297}\ (\bibinfo {organization} {American Institute of Physics},\
  \bibinfo {year} {2010})\ pp.\ \bibinfo {pages} {135--338}\BibitemShut
  {NoStop}%
\bibitem [{\citenamefont {Li}\ \emph {et~al.}(2015)\citenamefont {Li},
  \citenamefont {Jiang},\ and\ \citenamefont {Yao}}]{PhysRevB.91.241117}%
  \BibitemOpen
  \bibfield  {author} {\bibinfo {author} {\bibfnamefont {Z.-X.}\ \bibnamefont
  {Li}}, \bibinfo {author} {\bibfnamefont {Y.-F.}\ \bibnamefont {Jiang}},\ and\
  \bibinfo {author} {\bibfnamefont {H.}~\bibnamefont {Yao}},\ }\bibfield
  {title} {\bibinfo {title} {{Solving the fermion sign problem in quantum Monte
  Carlo simulations by Majorana representation}},\ }\href
  {https://doi.org/10.1103/PhysRevB.91.241117} {\bibfield  {journal} {\bibinfo
  {journal} {Phys. Rev. B}\ }\textbf {\bibinfo {volume} {91}},\ \bibinfo
  {pages} {241117} (\bibinfo {year} {2015})}\BibitemShut {NoStop}%
\bibitem [{\citenamefont {Wei}\ \emph {et~al.}(2016)\citenamefont {Wei},
  \citenamefont {Wu}, \citenamefont {Li}, \citenamefont {Zhang},\ and\
  \citenamefont {Xiang}}]{PhysRevLett.116.250601}%
  \BibitemOpen
  \bibfield  {author} {\bibinfo {author} {\bibfnamefont {Z.~C.}\ \bibnamefont
  {Wei}}, \bibinfo {author} {\bibfnamefont {C.}~\bibnamefont {Wu}}, \bibinfo
  {author} {\bibfnamefont {Y.}~\bibnamefont {Li}}, \bibinfo {author}
  {\bibfnamefont {S.}~\bibnamefont {Zhang}},\ and\ \bibinfo {author}
  {\bibfnamefont {T.}~\bibnamefont {Xiang}},\ }\bibfield  {title} {\bibinfo
  {title} {{Majorana Positivity and the Fermion Sign Problem of Quantum Monte
  Carlo Simulations}},\ }\href {https://doi.org/10.1103/PhysRevLett.116.250601}
  {\bibfield  {journal} {\bibinfo  {journal} {Phys. Rev. Lett.}\ }\textbf
  {\bibinfo {volume} {116}},\ \bibinfo {pages} {250601} (\bibinfo {year}
  {2016})}\BibitemShut {NoStop}%
\bibitem [{\citenamefont {Li}\ \emph {et~al.}(2016)\citenamefont {Li},
  \citenamefont {Jiang},\ and\ \citenamefont {Yao}}]{PhysRevLett.117.267002}%
  \BibitemOpen
  \bibfield  {author} {\bibinfo {author} {\bibfnamefont {Z.-X.}\ \bibnamefont
  {Li}}, \bibinfo {author} {\bibfnamefont {Y.-F.}\ \bibnamefont {Jiang}},\ and\
  \bibinfo {author} {\bibfnamefont {H.}~\bibnamefont {Yao}},\ }\bibfield
  {title} {\bibinfo {title} {{Majorana-Time-Reversal Symmetries: A Fundamental
  Principle for Sign-Problem-Free Quantum Monte Carlo Simulations}},\ }\href
  {https://doi.org/10.1103/PhysRevLett.117.267002} {\bibfield  {journal}
  {\bibinfo  {journal} {Phys. Rev. Lett.}\ }\textbf {\bibinfo {volume} {117}},\
  \bibinfo {pages} {267002} (\bibinfo {year} {2016})}\BibitemShut {NoStop}%
\bibitem [{\citenamefont {Wei}(2017)}]{wei2017semigroup}%
  \BibitemOpen
  \bibfield  {author} {\bibinfo {author} {\bibfnamefont {Z.-C.}\ \bibnamefont
  {Wei}},\ }\bibfield  {title} {\bibinfo {title} {{Semigroup Approach to the
  Sign Problem in Quantum Monte Carlo Simulations}},\ }\href@noop {} {\bibfield
   {journal} {\bibinfo  {journal} {arXiv preprint arXiv:1712.09412}\ }
  (\bibinfo {year} {2017})}\BibitemShut {NoStop}%
\bibitem [{\citenamefont {Sachdev}\ and\ \citenamefont
  {Ye}(1993)}]{Sachdev:1992fk}%
  \BibitemOpen
  \bibfield  {author} {\bibinfo {author} {\bibfnamefont {S.}~\bibnamefont
  {Sachdev}}\ and\ \bibinfo {author} {\bibfnamefont {J.}~\bibnamefont {Ye}},\
  }\bibfield  {title} {\bibinfo {title} {{Gapless spin-fluid ground state in a
  random quantum Heisenberg magnet}},\ }\href
  {https://doi.org/10.1103/PhysRevLett.70.3339} {\bibfield  {journal} {\bibinfo
   {journal} {Phys. Rev. Lett.}\ }\textbf {\bibinfo {volume} {70}},\ \bibinfo
  {pages} {3339} (\bibinfo {year} {1993})}\BibitemShut {NoStop}%
\bibitem [{\citenamefont {Kitaev}(2015{\natexlab{a}})}]{kitaevfirsttalk}%
  \BibitemOpen
  \bibfield  {author} {\bibinfo {author} {\bibfnamefont {A.}~\bibnamefont
  {Kitaev}},\ }\bibfield  {title} {\bibinfo {title} {{Hidden correlations in
  the Hawking radiation and thermal noise}},\ }\href@noop {} {\bibfield
  {journal} {\bibinfo  {journal}
  {\url{http://online.kitp.ucsb.edu/online/joint98/kitaev/}, KITP seminar, Feb.
  12}\ } (\bibinfo {year} {2015}{\natexlab{a}})}\BibitemShut {NoStop}%
\bibitem [{\citenamefont {Kitaev}(2015{\natexlab{b}})}]{KitaevTalks}%
  \BibitemOpen
  \bibfield  {author} {\bibinfo {author} {\bibfnamefont {A.}~\bibnamefont
  {Kitaev}},\ }\bibfield  {title} {\bibinfo {title} {{A simple model of quantum
  holography}},\ }\href@noop {} {\bibfield  {journal} {\bibinfo  {journal}
  {\url{http://online.kitp.ucsb.edu/online/entangled15/kitaev/},
  \url{http://online.kitp.ucsb.edu/online/entangled15/kitaev2/}, Talks at KITP,
  April 7 and May 27}\ } (\bibinfo {year} {2015}{\natexlab{b}})}\BibitemShut
  {NoStop}%
\bibitem [{\citenamefont {Polchinski}\ and\ \citenamefont
  {Rosenhaus}(2016)}]{Polchinski:2016xgd}%
  \BibitemOpen
  \bibfield  {author} {\bibinfo {author} {\bibfnamefont {J.}~\bibnamefont
  {Polchinski}}\ and\ \bibinfo {author} {\bibfnamefont {V.}~\bibnamefont
  {Rosenhaus}},\ }\bibfield  {title} {\bibinfo {title} {{The spectrum in the
  Sachdev-Ye-Kitaev model}},\ }\href {https://doi.org/10.1007/JHEP04(2016)001}
  {\bibfield  {journal} {\bibinfo  {journal} {Journal of High Energy Physics}\
  }\textbf {\bibinfo {volume} {2016}},\ \bibinfo {pages} {1} (\bibinfo {year}
  {2016})}\BibitemShut {NoStop}%
\bibitem [{\citenamefont {Jevicki}\ and\ \citenamefont
  {Suzuki}(2016{\natexlab{a}})}]{Jevicki:2016bwu}%
  \BibitemOpen
  \bibfield  {author} {\bibinfo {author} {\bibfnamefont {A.}~\bibnamefont
  {Jevicki}}\ and\ \bibinfo {author} {\bibfnamefont {K.}~\bibnamefont
  {Suzuki}},\ }\bibfield  {title} {\bibinfo {title} {{Bi-local holography in
  the SYK model: perturbations}},\ }\href
  {https://doi.org/10.1007/JHEP07(2016)007} {\bibfield  {journal} {\bibinfo
  {journal} {Journal of High Energy Physics}\ }\textbf {\bibinfo {volume}
  {2016}},\ \bibinfo {pages} {1} (\bibinfo {year}
  {2016}{\natexlab{a}})}\BibitemShut {NoStop}%
\bibitem [{\citenamefont {Maldacena}\ and\ \citenamefont
  {Stanford}(2016)}]{Maldacena:2016hyu}%
  \BibitemOpen
  \bibfield  {author} {\bibinfo {author} {\bibfnamefont {J.}~\bibnamefont
  {Maldacena}}\ and\ \bibinfo {author} {\bibfnamefont {D.}~\bibnamefont
  {Stanford}},\ }\bibfield  {title} {\bibinfo {title} {{Remarks on the
  Sachdev-Ye-Kitaev model}},\ }\href
  {https://doi.org/10.1103/PhysRevD.94.106002} {\bibfield  {journal} {\bibinfo
  {journal} {Phys. Rev. D}\ }\textbf {\bibinfo {volume} {94}},\ \bibinfo
  {pages} {106002} (\bibinfo {year} {2016})}\BibitemShut {NoStop}%
\bibitem [{\citenamefont {Jevicki}\ and\ \citenamefont
  {Suzuki}(2016{\natexlab{b}})}]{Jevicki:2016ito}%
  \BibitemOpen
  \bibfield  {author} {\bibinfo {author} {\bibfnamefont {A.}~\bibnamefont
  {Jevicki}}\ and\ \bibinfo {author} {\bibfnamefont {K.}~\bibnamefont
  {Suzuki}},\ }\bibfield  {title} {\bibinfo {title} {{Bi-local holography in
  the SYK model: perturbations}},\ }\href
  {https://doi.org/10.1007/JHEP11(2016)046} {\bibfield  {journal} {\bibinfo
  {journal} {Journal of High Energy Physics}\ }\textbf {\bibinfo {volume}
  {2016}},\ \bibinfo {pages} {1} (\bibinfo {year}
  {2016}{\natexlab{b}})}\BibitemShut {NoStop}%
\bibitem [{\citenamefont {Maldacena}\ \emph
  {et~al.}(2016{\natexlab{a}})\citenamefont {Maldacena}, \citenamefont
  {Shenker},\ and\ \citenamefont {Stanford}}]{Maldacena:2015waa}%
  \BibitemOpen
  \bibfield  {author} {\bibinfo {author} {\bibfnamefont {J.}~\bibnamefont
  {Maldacena}}, \bibinfo {author} {\bibfnamefont {S.~H.}\ \bibnamefont
  {Shenker}},\ and\ \bibinfo {author} {\bibfnamefont {D.}~\bibnamefont
  {Stanford}},\ }\bibfield  {title} {\bibinfo {title} {{A bound on chaos}},\
  }\href {https://doi.org/10.1007/JHEP08(2016)106} {\bibfield  {journal}
  {\bibinfo  {journal} {Journal of High Energy Physics}\ }\textbf {\bibinfo
  {volume} {2016}},\ \bibinfo {pages} {1} (\bibinfo {year}
  {2016}{\natexlab{a}})}\BibitemShut {NoStop}%
\bibitem [{\citenamefont {Roberts}\ \emph {et~al.}(2015)\citenamefont
  {Roberts}, \citenamefont {Stanford},\ and\ \citenamefont
  {Susskind}}]{Roberts:2014isa}%
  \BibitemOpen
  \bibfield  {author} {\bibinfo {author} {\bibfnamefont {D.~A.}\ \bibnamefont
  {Roberts}}, \bibinfo {author} {\bibfnamefont {D.}~\bibnamefont {Stanford}},\
  and\ \bibinfo {author} {\bibfnamefont {L.}~\bibnamefont {Susskind}},\
  }\bibfield  {title} {\bibinfo {title} {{Localized shocks}},\ }\href
  {https://doi.org/10.1007/JHEP03(2015)051} {\bibfield  {journal} {\bibinfo
  {journal} {Journal of High Energy Physics}\ }\textbf {\bibinfo {volume}
  {2015}},\ \bibinfo {pages} {1} (\bibinfo {year} {2015})}\BibitemShut
  {NoStop}%
\bibitem [{\citenamefont {Shenker}\ and\ \citenamefont
  {Stanford}(2015)}]{Shenker:2014cwa}%
  \BibitemOpen
  \bibfield  {author} {\bibinfo {author} {\bibfnamefont {S.~H.}\ \bibnamefont
  {Shenker}}\ and\ \bibinfo {author} {\bibfnamefont {D.}~\bibnamefont
  {Stanford}},\ }\bibfield  {title} {\bibinfo {title} {{Stringy effects in
  scrambling}},\ }\href {https://doi.org/10.1007/JHEP05(2015)132} {\bibfield
  {journal} {\bibinfo  {journal} {Journal of High Energy Physics}\ }\textbf
  {\bibinfo {volume} {2015}},\ \bibinfo {pages} {1} (\bibinfo {year}
  {2015})}\BibitemShut {NoStop}%
\bibitem [{\citenamefont {Maldacena}\ \emph
  {et~al.}(2016{\natexlab{b}})\citenamefont {Maldacena}, \citenamefont
  {Stanford},\ and\ \citenamefont {Yang}}]{Maldacena:2016upp}%
  \BibitemOpen
  \bibfield  {author} {\bibinfo {author} {\bibfnamefont {J.}~\bibnamefont
  {Maldacena}}, \bibinfo {author} {\bibfnamefont {D.}~\bibnamefont
  {Stanford}},\ and\ \bibinfo {author} {\bibfnamefont {Z.}~\bibnamefont
  {Yang}},\ }\bibfield  {title} {\bibinfo {title} {{Conformal symmetry and its
  breaking in two-dimensional nearly anti-de Sitter space}},\ }\bibfield
  {journal} {\bibinfo  {journal} {Progress of Theoretical and Experimental
  Physics}\ }\textbf {\bibinfo {volume} {2016}},\ \href
  {https://doi.org/10.1093/ptep/ptw124} {10.1093/ptep/ptw124} (\bibinfo {year}
  {2016}{\natexlab{b}})\BibitemShut {NoStop}%
\bibitem [{\citenamefont {Fu}\ \emph {et~al.}(2017)\citenamefont {Fu},
  \citenamefont {Gaiotto}, \citenamefont {Maldacena},\ and\ \citenamefont
  {Sachdev}}]{Fu:2016vas}%
  \BibitemOpen
  \bibfield  {author} {\bibinfo {author} {\bibfnamefont {W.}~\bibnamefont
  {Fu}}, \bibinfo {author} {\bibfnamefont {D.}~\bibnamefont {Gaiotto}},
  \bibinfo {author} {\bibfnamefont {J.}~\bibnamefont {Maldacena}},\ and\
  \bibinfo {author} {\bibfnamefont {S.}~\bibnamefont {Sachdev}},\ }\bibfield
  {title} {\bibinfo {title} {{Supersymmetric Sachdev-Ye-Kitaev models}},\
  }\href {https://doi.org/10.1103/PhysRevD.95.026009} {\bibfield  {journal}
  {\bibinfo  {journal} {Phys. Rev. D}\ }\textbf {\bibinfo {volume} {95}},\
  \bibinfo {pages} {026009} (\bibinfo {year} {2017})},\ \bibinfo {note}
  {[Erratum: Phys.Rev.D 95, 069904 (2017)]}\BibitemShut {NoStop}%
\bibitem [{\citenamefont {Cotler}\ \emph {et~al.}(2017)\citenamefont {Cotler},
  \citenamefont {Gur-Ari}, \citenamefont {Hanada}, \citenamefont {Polchinski},
  \citenamefont {Saad}, \citenamefont {Shenker}, \citenamefont {Stanford},
  \citenamefont {Streicher},\ and\ \citenamefont {Tezuka}}]{Cotler:2016fpe}%
  \BibitemOpen
  \bibfield  {author} {\bibinfo {author} {\bibfnamefont {J.~S.}\ \bibnamefont
  {Cotler}}, \bibinfo {author} {\bibfnamefont {G.}~\bibnamefont {Gur-Ari}},
  \bibinfo {author} {\bibfnamefont {M.}~\bibnamefont {Hanada}}, \bibinfo
  {author} {\bibfnamefont {J.}~\bibnamefont {Polchinski}}, \bibinfo {author}
  {\bibfnamefont {P.}~\bibnamefont {Saad}}, \bibinfo {author} {\bibfnamefont
  {S.~H.}\ \bibnamefont {Shenker}}, \bibinfo {author} {\bibfnamefont
  {D.}~\bibnamefont {Stanford}}, \bibinfo {author} {\bibfnamefont
  {A.}~\bibnamefont {Streicher}},\ and\ \bibinfo {author} {\bibfnamefont
  {M.}~\bibnamefont {Tezuka}},\ }\bibfield  {title} {\bibinfo {title} {{Black
  holes and random matrices}},\ }\href
  {https://doi.org/10.1007/JHEP05(2017)118} {\bibfield  {journal} {\bibinfo
  {journal} {Journal of High Energy Physics}\ }\textbf {\bibinfo {volume}
  {2017}},\ \bibinfo {pages} {1} (\bibinfo {year} {2017})},\ \bibinfo {note}
  {[Erratum: JHEP 09, 002 (2018)]}\BibitemShut {NoStop}%
\bibitem [{\citenamefont {Yoon}(2017)}]{Yoon:2017nig}%
  \BibitemOpen
  \bibfield  {author} {\bibinfo {author} {\bibfnamefont {J.}~\bibnamefont
  {Yoon}},\ }\bibfield  {title} {\bibinfo {title} {{SYK models and SYK-like
  tensor models with global symmetry}},\ }\href
  {https://doi.org/10.1007/JHEP10(2017)183} {\bibfield  {journal} {\bibinfo
  {journal} {Journal of High Energy Physics}\ }\textbf {\bibinfo {volume}
  {2017}},\ \bibinfo {pages} {1} (\bibinfo {year} {2017})}\BibitemShut
  {NoStop}%
\bibitem [{\citenamefont {Li}\ \emph {et~al.}(2017)\citenamefont {Li},
  \citenamefont {Liu}, \citenamefont {Xin},\ and\ \citenamefont
  {Zhou}}]{Li:2017hdt}%
  \BibitemOpen
  \bibfield  {author} {\bibinfo {author} {\bibfnamefont {T.}~\bibnamefont
  {Li}}, \bibinfo {author} {\bibfnamefont {J.}~\bibnamefont {Liu}}, \bibinfo
  {author} {\bibfnamefont {Y.}~\bibnamefont {Xin}},\ and\ \bibinfo {author}
  {\bibfnamefont {Y.}~\bibnamefont {Zhou}},\ }\bibfield  {title} {\bibinfo
  {title} {{Supersymmetric SYK model and random matrix theory}},\ }\href
  {https://doi.org/10.1007/JHEP06(2017)111} {\bibfield  {journal} {\bibinfo
  {journal} {Journal of High Energy Physics}\ }\textbf {\bibinfo {volume}
  {2017}},\ \bibinfo {pages} {1} (\bibinfo {year} {2017})}\BibitemShut
  {NoStop}%
\bibitem [{\citenamefont {Banerjee}\ and\ \citenamefont
  {Altman}(2017)}]{PhysRevB.95.134302}%
  \BibitemOpen
  \bibfield  {author} {\bibinfo {author} {\bibfnamefont {S.}~\bibnamefont
  {Banerjee}}\ and\ \bibinfo {author} {\bibfnamefont {E.}~\bibnamefont
  {Altman}},\ }\bibfield  {title} {\bibinfo {title} {{Solvable model for a
  dynamical quantum phase transition from fast to slow scrambling}},\ }\href
  {https://doi.org/10.1103/PhysRevB.95.134302} {\bibfield  {journal} {\bibinfo
  {journal} {Phys. Rev. B}\ }\textbf {\bibinfo {volume} {95}},\ \bibinfo
  {pages} {134302} (\bibinfo {year} {2017})}\BibitemShut {NoStop}%
\bibitem [{\citenamefont {Garc\'{\i}a-Garc\'{\i}a}\ \emph
  {et~al.}(2018)\citenamefont {Garc\'{\i}a-Garc\'{\i}a}, \citenamefont
  {Loureiro}, \citenamefont {Romero-Berm\'udez},\ and\ \citenamefont
  {Tezuka}}]{Garcia-Garcia:2017bkg}%
  \BibitemOpen
  \bibfield  {author} {\bibinfo {author} {\bibfnamefont {A.~M.}\ \bibnamefont
  {Garc\'{\i}a-Garc\'{\i}a}}, \bibinfo {author} {\bibfnamefont
  {B.}~\bibnamefont {Loureiro}}, \bibinfo {author} {\bibfnamefont
  {A.}~\bibnamefont {Romero-Berm\'udez}},\ and\ \bibinfo {author}
  {\bibfnamefont {M.}~\bibnamefont {Tezuka}},\ }\bibfield  {title} {\bibinfo
  {title} {{Chaotic-Integrable Transition in the Sachdev-Ye-Kitaev Model}},\
  }\href {https://doi.org/10.1103/PhysRevLett.120.241603} {\bibfield  {journal}
  {\bibinfo  {journal} {Phys. Rev. Lett.}\ }\textbf {\bibinfo {volume} {120}},\
  \bibinfo {pages} {241603} (\bibinfo {year} {2018})}\BibitemShut {NoStop}%
\bibitem [{\citenamefont {Pan}\ \emph {et~al.}(2021)\citenamefont {Pan},
  \citenamefont {Wang}, \citenamefont {Davis}, \citenamefont {Wang},\ and\
  \citenamefont {Meng}}]{pan2020self}%
  \BibitemOpen
  \bibfield  {author} {\bibinfo {author} {\bibfnamefont {G.}~\bibnamefont
  {Pan}}, \bibinfo {author} {\bibfnamefont {W.}~\bibnamefont {Wang}}, \bibinfo
  {author} {\bibfnamefont {A.}~\bibnamefont {Davis}}, \bibinfo {author}
  {\bibfnamefont {Y.}~\bibnamefont {Wang}},\ and\ \bibinfo {author}
  {\bibfnamefont {Z.~Y.}\ \bibnamefont {Meng}},\ }\bibfield  {title} {\bibinfo
  {title} {{Yukawa-SYK model and self-tuned quantum criticality}},\ }\href
  {https://doi.org/10.1103/PhysRevResearch.3.013250} {\bibfield  {journal}
  {\bibinfo  {journal} {Phys. Rev. Research}\ }\textbf {\bibinfo {volume}
  {3}},\ \bibinfo {pages} {013250} (\bibinfo {year} {2021})}\BibitemShut
  {NoStop}%
\bibitem [{\citenamefont {Wang}\ \emph {et~al.}(2021)\citenamefont {Wang},
  \citenamefont {Davis}, \citenamefont {Pan}, \citenamefont {Wang},\ and\
  \citenamefont {Meng}}]{PhysRevB.103.195108}%
  \BibitemOpen
  \bibfield  {author} {\bibinfo {author} {\bibfnamefont {W.}~\bibnamefont
  {Wang}}, \bibinfo {author} {\bibfnamefont {A.}~\bibnamefont {Davis}},
  \bibinfo {author} {\bibfnamefont {G.}~\bibnamefont {Pan}}, \bibinfo {author}
  {\bibfnamefont {Y.}~\bibnamefont {Wang}},\ and\ \bibinfo {author}
  {\bibfnamefont {Z.~Y.}\ \bibnamefont {Meng}},\ }\bibfield  {title} {\bibinfo
  {title} {{Phase diagram of the spin-$\frac{1}{2}$ Yukawa--Sachdev-Ye-Kitaev
  model: Non-Fermi liquid, insulator, and superconductor}},\ }\href
  {https://doi.org/10.1103/PhysRevB.103.195108} {\bibfield  {journal} {\bibinfo
   {journal} {Phys. Rev. B}\ }\textbf {\bibinfo {volume} {103}},\ \bibinfo
  {pages} {195108} (\bibinfo {year} {2021})}\BibitemShut {NoStop}%
\bibitem [{\citenamefont {Haldar}\ \emph {et~al.}(2020)\citenamefont {Haldar},
  \citenamefont {Bera},\ and\ \citenamefont
  {Banerjee}}]{PhysRevResearch.2.033505}%
  \BibitemOpen
  \bibfield  {author} {\bibinfo {author} {\bibfnamefont {A.}~\bibnamefont
  {Haldar}}, \bibinfo {author} {\bibfnamefont {S.}~\bibnamefont {Bera}},\ and\
  \bibinfo {author} {\bibfnamefont {S.}~\bibnamefont {Banerjee}},\ }\bibfield
  {title} {\bibinfo {title} {{R\'enyi entanglement entropy of Fermi and
  non-Fermi liquids: Sachdev-Ye-Kitaev model and dynamical mean field
  theories}},\ }\href {https://doi.org/10.1103/PhysRevResearch.2.033505}
  {\bibfield  {journal} {\bibinfo  {journal} {Phys. Rev. Research}\ }\textbf
  {\bibinfo {volume} {2}},\ \bibinfo {pages} {033505} (\bibinfo {year}
  {2020})}\BibitemShut {NoStop}%
\bibitem [{\citenamefont {Sahoo}\ \emph {et~al.}(2020)\citenamefont {Sahoo},
  \citenamefont {Lantagne-Hurtubise}, \citenamefont {Plugge},\ and\
  \citenamefont {Franz}}]{PhysRevResearch.2.043049}%
  \BibitemOpen
  \bibfield  {author} {\bibinfo {author} {\bibfnamefont {S.}~\bibnamefont
  {Sahoo}}, \bibinfo {author} {\bibfnamefont {E.}~\bibnamefont
  {Lantagne-Hurtubise}}, \bibinfo {author} {\bibfnamefont {S.}~\bibnamefont
  {Plugge}},\ and\ \bibinfo {author} {\bibfnamefont {M.}~\bibnamefont
  {Franz}},\ }\bibfield  {title} {\bibinfo {title} {{Traversable wormhole and
  Hawking-Page transition in coupled complex SYK models}},\ }\href
  {https://doi.org/10.1103/PhysRevResearch.2.043049} {\bibfield  {journal}
  {\bibinfo  {journal} {Phys. Rev. Research}\ }\textbf {\bibinfo {volume}
  {2}},\ \bibinfo {pages} {043049} (\bibinfo {year} {2020})}\BibitemShut
  {NoStop}%
\bibitem [{\citenamefont {Lantagne-Hurtubise}\ \emph
  {et~al.}(2021)\citenamefont {Lantagne-Hurtubise}, \citenamefont {Pathak},
  \citenamefont {Sahoo},\ and\ \citenamefont
  {Franz}}]{lantagne2020superconducting}%
  \BibitemOpen
  \bibfield  {author} {\bibinfo {author} {\bibfnamefont {E.}~\bibnamefont
  {Lantagne-Hurtubise}}, \bibinfo {author} {\bibfnamefont {V.}~\bibnamefont
  {Pathak}}, \bibinfo {author} {\bibfnamefont {S.}~\bibnamefont {Sahoo}},\ and\
  \bibinfo {author} {\bibfnamefont {M.}~\bibnamefont {Franz}},\ }\bibfield
  {title} {\bibinfo {title} {{Superconducting instabilities in a spinful
  Sachdev-Ye-Kitaev model}},\ }\href
  {https://doi.org/10.1103/PhysRevB.104.L020509} {\bibfield  {journal}
  {\bibinfo  {journal} {Phys. Rev. B}\ }\textbf {\bibinfo {volume} {104}},\
  \bibinfo {pages} {L020509} (\bibinfo {year} {2021})}\BibitemShut {NoStop}%
\bibitem [{\citenamefont {Haenel}\ \emph {et~al.}()\citenamefont {Haenel},
  \citenamefont {Sahoo}, \citenamefont {Hsieh},\ and\ \citenamefont
  {Franz}}]{haenel2021traversable}%
  \BibitemOpen
  \bibfield  {author} {\bibinfo {author} {\bibfnamefont {R.}~\bibnamefont
  {Haenel}}, \bibinfo {author} {\bibfnamefont {S.}~\bibnamefont {Sahoo}},
  \bibinfo {author} {\bibfnamefont {T.~H.}\ \bibnamefont {Hsieh}},\ and\
  \bibinfo {author} {\bibfnamefont {M.}~\bibnamefont {Franz}},\ }\bibfield
  {title} {\bibinfo {title} {{Traversable wormhole in coupled SYK models with
  imbalanced interactions}},\ }\href@noop {} {\ }\Eprint
  {https://arxiv.org/abs/2102.05687} {arXiv:2102.05687} \BibitemShut {NoStop}%
\bibitem [{\citenamefont {You}\ \emph {et~al.}(2017)\citenamefont {You},
  \citenamefont {Ludwig},\ and\ \citenamefont {Xu}}]{PhysRevB.95.115150}%
  \BibitemOpen
  \bibfield  {author} {\bibinfo {author} {\bibfnamefont {Y.-Z.}\ \bibnamefont
  {You}}, \bibinfo {author} {\bibfnamefont {A.~W.~W.}\ \bibnamefont {Ludwig}},\
  and\ \bibinfo {author} {\bibfnamefont {C.}~\bibnamefont {Xu}},\ }\bibfield
  {title} {\bibinfo {title} {{Sachdev-Ye-Kitaev model and thermalization on the
  boundary of many-body localized fermionic symmetry-protected topological
  states}},\ }\href {https://doi.org/10.1103/PhysRevB.95.115150} {\bibfield
  {journal} {\bibinfo  {journal} {Phys. Rev. B}\ }\textbf {\bibinfo {volume}
  {95}},\ \bibinfo {pages} {115150} (\bibinfo {year} {2017})}\BibitemShut
  {NoStop}%
\bibitem [{\citenamefont {Atas}\ \emph {et~al.}(2013)\citenamefont {Atas},
  \citenamefont {Bogomolny}, \citenamefont {Giraud},\ and\ \citenamefont
  {Roux}}]{PhysRevLett.110.084101}%
  \BibitemOpen
  \bibfield  {author} {\bibinfo {author} {\bibfnamefont {Y.~Y.}\ \bibnamefont
  {Atas}}, \bibinfo {author} {\bibfnamefont {E.}~\bibnamefont {Bogomolny}},
  \bibinfo {author} {\bibfnamefont {O.}~\bibnamefont {Giraud}},\ and\ \bibinfo
  {author} {\bibfnamefont {G.}~\bibnamefont {Roux}},\ }\bibfield  {title}
  {\bibinfo {title} {{Distribution of the Ratio of Consecutive Level Spacings
  in Random Matrix Ensembles}},\ }\href
  {https://doi.org/10.1103/PhysRevLett.110.084101} {\bibfield  {journal}
  {\bibinfo  {journal} {Phys. Rev. Lett.}\ }\textbf {\bibinfo {volume} {110}},\
  \bibinfo {pages} {084101} (\bibinfo {year} {2013})}\BibitemShut {NoStop}%
\bibitem [{\citenamefont {Zirnbauer}(1996)}]{zirnbauer1996riemannian}%
  \BibitemOpen
  \bibfield  {author} {\bibinfo {author} {\bibfnamefont {M.~R.}\ \bibnamefont
  {Zirnbauer}},\ }\bibfield  {title} {\bibinfo {title} {{Riemannian symmetric
  superspaces and their origin in random-matrix theory}},\ }\href
  {https://doi.org/10.1063/1.531675} {\bibfield  {journal} {\bibinfo  {journal}
  {Journal of Mathematical Physics}\ }\textbf {\bibinfo {volume} {37}},\
  \bibinfo {pages} {4986} (\bibinfo {year} {1996})}\BibitemShut {NoStop}%
\bibitem [{\citenamefont {Altland}\ and\ \citenamefont
  {Zirnbauer}(1997)}]{PhysRevB.55.1142}%
  \BibitemOpen
  \bibfield  {author} {\bibinfo {author} {\bibfnamefont {A.}~\bibnamefont
  {Altland}}\ and\ \bibinfo {author} {\bibfnamefont {M.~R.}\ \bibnamefont
  {Zirnbauer}},\ }\bibfield  {title} {\bibinfo {title} {{Nonstandard symmetry
  classes in mesoscopic normal-superconducting hybrid structures}},\ }\href
  {https://doi.org/10.1103/PhysRevB.55.1142} {\bibfield  {journal} {\bibinfo
  {journal} {Phys. Rev. B}\ }\textbf {\bibinfo {volume} {55}},\ \bibinfo
  {pages} {1142} (\bibinfo {year} {1997})}\BibitemShut {NoStop}%
\bibitem [{\citenamefont {Binder}\ and\ \citenamefont
  {Young}(1986)}]{RevModPhys.58.801}%
  \BibitemOpen
  \bibfield  {author} {\bibinfo {author} {\bibfnamefont {K.}~\bibnamefont
  {Binder}}\ and\ \bibinfo {author} {\bibfnamefont {A.~P.}\ \bibnamefont
  {Young}},\ }\bibfield  {title} {\bibinfo {title} {{Spin glasses: Experimental
  facts, theoretical concepts, and open questions}},\ }\href
  {https://doi.org/10.1103/RevModPhys.58.801} {\bibfield  {journal} {\bibinfo
  {journal} {Rev. Mod. Phys.}\ }\textbf {\bibinfo {volume} {58}},\ \bibinfo
  {pages} {801} (\bibinfo {year} {1986})}\BibitemShut {NoStop}%
\bibitem [{\citenamefont {Kang}\ \emph {et~al.}(2020)\citenamefont {Kang},
  \citenamefont {Parameswaran}, \citenamefont {Potter}, \citenamefont
  {Vasseur},\ and\ \citenamefont {Gazit}}]{PhysRevB.102.224204}%
  \BibitemOpen
  \bibfield  {author} {\bibinfo {author} {\bibfnamefont {B.}~\bibnamefont
  {Kang}}, \bibinfo {author} {\bibfnamefont {S.~A.}\ \bibnamefont
  {Parameswaran}}, \bibinfo {author} {\bibfnamefont {A.~C.}\ \bibnamefont
  {Potter}}, \bibinfo {author} {\bibfnamefont {R.}~\bibnamefont {Vasseur}},\
  and\ \bibinfo {author} {\bibfnamefont {S.}~\bibnamefont {Gazit}},\ }\bibfield
   {title} {\bibinfo {title} {{Superuniversality from disorder at
  two-dimensional topological phase transitions}},\ }\href
  {https://doi.org/10.1103/PhysRevB.102.224204} {\bibfield  {journal} {\bibinfo
   {journal} {Phys. Rev. B}\ }\textbf {\bibinfo {volume} {102}},\ \bibinfo
  {pages} {224204} (\bibinfo {year} {2020})}\BibitemShut {NoStop}%
\bibitem [{\citenamefont {Liu}\ \emph {et~al.}(2018)\citenamefont {Liu},
  \citenamefont {Chen},\ and\ \citenamefont {Balents}}]{PhysRevB.97.245126}%
  \BibitemOpen
  \bibfield  {author} {\bibinfo {author} {\bibfnamefont {C.}~\bibnamefont
  {Liu}}, \bibinfo {author} {\bibfnamefont {X.}~\bibnamefont {Chen}},\ and\
  \bibinfo {author} {\bibfnamefont {L.}~\bibnamefont {Balents}},\ }\bibfield
  {title} {\bibinfo {title} {{Quantum entanglement of the Sachdev-Ye-Kitaev
  models}},\ }\href {https://doi.org/10.1103/PhysRevB.97.245126} {\bibfield
  {journal} {\bibinfo  {journal} {Phys. Rev. B}\ }\textbf {\bibinfo {volume}
  {97}},\ \bibinfo {pages} {245126} (\bibinfo {year} {2018})}\BibitemShut
  {NoStop}%
\bibitem [{\citenamefont {Huang}\ and\ \citenamefont
  {Gu}(2019)}]{PhysRevD.100.041901}%
  \BibitemOpen
  \bibfield  {author} {\bibinfo {author} {\bibfnamefont {Y.}~\bibnamefont
  {Huang}}\ and\ \bibinfo {author} {\bibfnamefont {Y.}~\bibnamefont {Gu}},\
  }\bibfield  {title} {\bibinfo {title} {{Eigenstate entanglement in the
  Sachdev-Ye-Kitaev model}},\ }\href
  {https://doi.org/10.1103/PhysRevD.100.041901} {\bibfield  {journal} {\bibinfo
   {journal} {Phys. Rev. D}\ }\textbf {\bibinfo {volume} {100}},\ \bibinfo
  {pages} {041901} (\bibinfo {year} {2019})}\BibitemShut {NoStop}%
\bibitem [{\citenamefont {Zhang}\ \emph {et~al.}(2020)\citenamefont {Zhang},
  \citenamefont {Liu},\ and\ \citenamefont
  {Chen}}]{10.21468/SciPostPhys.8.6.094}%
  \BibitemOpen
  \bibfield  {author} {\bibinfo {author} {\bibfnamefont {P.}~\bibnamefont
  {Zhang}}, \bibinfo {author} {\bibfnamefont {C.}~\bibnamefont {Liu}},\ and\
  \bibinfo {author} {\bibfnamefont {X.}~\bibnamefont {Chen}},\ }\bibfield
  {title} {\bibinfo {title} {{Subsystem R{\'e}nyi Entropy of Thermal Ensembles
  for SYK-like models}},\ }\href {https://doi.org/10.21468/SciPostPhys.8.6.094}
  {\bibfield  {journal} {\bibinfo  {journal} {SciPost Phys.}\ }\textbf
  {\bibinfo {volume} {8}},\ \bibinfo {pages} {94} (\bibinfo {year}
  {2020})}\BibitemShut {NoStop}%
\bibitem [{\citenamefont {Maldacena}\ and\ \citenamefont
  {Qi}()}]{maldacena2018eternal}%
  \BibitemOpen
  \bibfield  {author} {\bibinfo {author} {\bibfnamefont {J.}~\bibnamefont
  {Maldacena}}\ and\ \bibinfo {author} {\bibfnamefont {X.-L.}\ \bibnamefont
  {Qi}},\ }\bibfield  {title} {\bibinfo {title} {{Eternal traversable
  wormhole}},\ }\href@noop {} {\ }\Eprint {https://arxiv.org/abs/1804.00491}
  {arXiv:1804.00491} \BibitemShut {NoStop}%
\bibitem [{\citenamefont {Witten}(2019)}]{Witten:2016iux}%
  \BibitemOpen
  \bibfield  {author} {\bibinfo {author} {\bibfnamefont {E.}~\bibnamefont
  {Witten}},\ }\bibfield  {title} {\bibinfo {title} {{An SYK-like model without
  disorder}},\ }\href {https://doi.org/10.1088/1751-8121/ab3752} {\bibfield
  {journal} {\bibinfo  {journal} {Journal of Physics A: Mathematical and
  Theoretical}\ }\textbf {\bibinfo {volume} {52}},\ \bibinfo {pages} {474002}
  (\bibinfo {year} {2019})}\BibitemShut {NoStop}%
\bibitem [{\citenamefont {Gurau}(2017)}]{Gurau:2016lzk}%
  \BibitemOpen
  \bibfield  {author} {\bibinfo {author} {\bibfnamefont {R.}~\bibnamefont
  {Gurau}},\ }\bibfield  {title} {\bibinfo {title} {{The complete $1/N$
  expansion of a SYK--like tensor model}},\ }\href
  {https://doi.org/10.1016/j.nuclphysb.2017.01.015} {\bibfield  {journal}
  {\bibinfo  {journal} {Nuclear Physics B}\ }\textbf {\bibinfo {volume}
  {916}},\ \bibinfo {pages} {386} (\bibinfo {year} {2017})}\BibitemShut
  {NoStop}%
\bibitem [{\citenamefont {Klebanov}\ and\ \citenamefont
  {Tarnopolsky}(2017)}]{Klebanov:2016xxf}%
  \BibitemOpen
  \bibfield  {author} {\bibinfo {author} {\bibfnamefont {I.~R.}\ \bibnamefont
  {Klebanov}}\ and\ \bibinfo {author} {\bibfnamefont {G.}~\bibnamefont
  {Tarnopolsky}},\ }\bibfield  {title} {\bibinfo {title} {{Uncolored random
  tensors, melon diagrams, and the Sachdev-Ye-Kitaev models}},\ }\href
  {https://doi.org/10.1103/PhysRevD.95.046004} {\bibfield  {journal} {\bibinfo
  {journal} {Phys. Rev. D}\ }\textbf {\bibinfo {volume} {95}},\ \bibinfo
  {pages} {046004} (\bibinfo {year} {2017})}\BibitemShut {NoStop}%
\bibitem [{\citenamefont {Narayan}\ and\ \citenamefont
  {Yoon}(2017)}]{Narayan:2017qtw}%
  \BibitemOpen
  \bibfield  {author} {\bibinfo {author} {\bibfnamefont {P.}~\bibnamefont
  {Narayan}}\ and\ \bibinfo {author} {\bibfnamefont {J.}~\bibnamefont {Yoon}},\
  }\bibfield  {title} {\bibinfo {title} {{SYK-like tensor models on the
  lattice}},\ }\href {https://doi.org/10.1007/JHEP08(2017)083} {\bibfield
  {journal} {\bibinfo  {journal} {Journal of High Energy Physics}\ }\textbf
  {\bibinfo {volume} {2017}},\ \bibinfo {pages} {1} (\bibinfo {year}
  {2017})}\BibitemShut {NoStop}%
\bibitem [{\citenamefont {Giombi}\ \emph {et~al.}(2018)\citenamefont {Giombi},
  \citenamefont {Klebanov}, \citenamefont {Popov}, \citenamefont {Prakash},\
  and\ \citenamefont {Tarnopolsky}}]{Giombi:2018qgp}%
  \BibitemOpen
  \bibfield  {author} {\bibinfo {author} {\bibfnamefont {S.}~\bibnamefont
  {Giombi}}, \bibinfo {author} {\bibfnamefont {I.~R.}\ \bibnamefont
  {Klebanov}}, \bibinfo {author} {\bibfnamefont {F.}~\bibnamefont {Popov}},
  \bibinfo {author} {\bibfnamefont {S.}~\bibnamefont {Prakash}},\ and\ \bibinfo
  {author} {\bibfnamefont {G.}~\bibnamefont {Tarnopolsky}},\ }\bibfield
  {title} {\bibinfo {title} {{Prismatic large $N$ models for bosonic
  tensors}},\ }\href {https://doi.org/10.1103/PhysRevD.98.105005} {\bibfield
  {journal} {\bibinfo  {journal} {Phys. Rev. D}\ }\textbf {\bibinfo {volume}
  {98}},\ \bibinfo {pages} {105005} (\bibinfo {year} {2018})}\BibitemShut
  {NoStop}%
\bibitem [{\citenamefont {de~Mello~Koch}\ \emph {et~al.}(2017)\citenamefont
  {de~Mello~Koch}, \citenamefont {Gossman},\ and\ \citenamefont
  {Tribelhorn}}]{deMelloKoch:2017bvv}%
  \BibitemOpen
  \bibfield  {author} {\bibinfo {author} {\bibfnamefont {R.}~\bibnamefont
  {de~Mello~Koch}}, \bibinfo {author} {\bibfnamefont {D.}~\bibnamefont
  {Gossman}},\ and\ \bibinfo {author} {\bibfnamefont {L.}~\bibnamefont
  {Tribelhorn}},\ }\bibfield  {title} {\bibinfo {title} {{Gauge invariants,
  correlators and holography in bosonic and fermionic tensor models}},\ }\href
  {https://doi.org/10.1007/JHEP09(2017)011} {\bibfield  {journal} {\bibinfo
  {journal} {Journal of High Energy Physics}\ }\textbf {\bibinfo {volume}
  {2017}},\ \bibinfo {pages} {1} (\bibinfo {year} {2017})}\BibitemShut
  {NoStop}%
\bibitem [{\citenamefont {Azeyanagi}\ \emph {et~al.}(2018)\citenamefont
  {Azeyanagi}, \citenamefont {Ferrari},\ and\ \citenamefont
  {Massolo}}]{Azeyanagi:2017drg}%
  \BibitemOpen
  \bibfield  {author} {\bibinfo {author} {\bibfnamefont {T.}~\bibnamefont
  {Azeyanagi}}, \bibinfo {author} {\bibfnamefont {F.}~\bibnamefont {Ferrari}},\
  and\ \bibinfo {author} {\bibfnamefont {F.~I.~S.}\ \bibnamefont {Massolo}},\
  }\bibfield  {title} {\bibinfo {title} {{Phase Diagram of Planar Matrix
  Quantum Mechanics, Tensor, and Sachdev-Ye-Kitaev Models}},\ }\href
  {https://doi.org/10.1103/PhysRevLett.120.061602} {\bibfield  {journal}
  {\bibinfo  {journal} {Phys. Rev. Lett.}\ }\textbf {\bibinfo {volume} {120}},\
  \bibinfo {pages} {061602} (\bibinfo {year} {2018})}\BibitemShut {NoStop}%
\bibitem [{\citenamefont {Ferrari}\ and\ \citenamefont
  {Schaposnik~Massolo}(2019)}]{Ferrari:2019ogc}%
  \BibitemOpen
  \bibfield  {author} {\bibinfo {author} {\bibfnamefont {F.}~\bibnamefont
  {Ferrari}}\ and\ \bibinfo {author} {\bibfnamefont {F.~I.}\ \bibnamefont
  {Schaposnik~Massolo}},\ }\bibfield  {title} {\bibinfo {title} {{Phases of
  melonic quantum mechanics}},\ }\href
  {https://doi.org/10.1103/PhysRevD.100.026007} {\bibfield  {journal} {\bibinfo
   {journal} {Phys. Rev. D}\ }\textbf {\bibinfo {volume} {100}},\ \bibinfo
  {pages} {026007} (\bibinfo {year} {2019})}\BibitemShut {NoStop}%
\bibitem [{\citenamefont {$\textrm{P. Saad}$}\ \emph {et~al.}()\citenamefont
  {$\textrm{P. Saad}$}, \citenamefont {Shenker},\ and\ \citenamefont
  {Stanford}}]{saad2018semiclassical}%
  \BibitemOpen
  \bibfield  {author} {\bibinfo {author} {\bibnamefont {$\textrm{P. Saad}$}},
  \bibinfo {author} {\bibfnamefont {S.~H.}\ \bibnamefont {Shenker}},\ and\
  \bibinfo {author} {\bibfnamefont {D.}~\bibnamefont {Stanford}},\ }\bibfield
  {title} {\bibinfo {title} {{A semiclassical ramp in SYK and in gravity}},\
  }\href@noop {} {\ }\Eprint {https://arxiv.org/abs/1806.06840}
  {arXiv:1806.06840} \BibitemShut {NoStop}%
\bibitem [{\citenamefont {Saad}\ \emph {et~al.}()\citenamefont {Saad},
  \citenamefont {Shenker},\ and\ \citenamefont {Stanford}}]{saad2019jt}%
  \BibitemOpen
  \bibfield  {author} {\bibinfo {author} {\bibfnamefont {P.}~\bibnamefont
  {Saad}}, \bibinfo {author} {\bibfnamefont {S.~H.}\ \bibnamefont {Shenker}},\
  and\ \bibinfo {author} {\bibfnamefont {D.}~\bibnamefont {Stanford}},\
  }\bibfield  {title} {\bibinfo {title} {{JT gravity as a matrix integral}},\
  }\href@noop {} {\ }\Eprint {https://arxiv.org/abs/1903.11115}
  {arXiv:1903.11115} \BibitemShut {NoStop}%
\bibitem [{\citenamefont {Stanford}\ and\ \citenamefont
  {Witten}()}]{Stanford:2019vob}%
  \BibitemOpen
  \bibfield  {author} {\bibinfo {author} {\bibfnamefont {D.}~\bibnamefont
  {Stanford}}\ and\ \bibinfo {author} {\bibfnamefont {E.}~\bibnamefont
  {Witten}},\ }\bibfield  {title} {\bibinfo {title} {{JT Gravity and the
  Ensembles of Random Matrix Theory}},\ }\href@noop {} {\ }\Eprint
  {https://arxiv.org/abs/1907.03363} {arXiv:1907.03363} \BibitemShut {NoStop}%
\bibitem [{\citenamefont {Bohigas}\ \emph {et~al.}(1984)\citenamefont
  {Bohigas}, \citenamefont {Giannoni},\ and\ \citenamefont
  {Schmit}}]{Bohigas:1983er}%
  \BibitemOpen
  \bibfield  {author} {\bibinfo {author} {\bibfnamefont {O.}~\bibnamefont
  {Bohigas}}, \bibinfo {author} {\bibfnamefont {M.~J.}\ \bibnamefont
  {Giannoni}},\ and\ \bibinfo {author} {\bibfnamefont {C.}~\bibnamefont
  {Schmit}},\ }\bibfield  {title} {\bibinfo {title} {{Characterization of
  chaotic quantum spectra and universality of level fluctuation laws}},\ }\href
  {https://doi.org/10.1103/PhysRevLett.52.1} {\bibfield  {journal} {\bibinfo
  {journal} {Phys. Rev. Lett.}\ }\textbf {\bibinfo {volume} {52}},\ \bibinfo
  {pages} {1} (\bibinfo {year} {1984})}\BibitemShut {NoStop}%
\bibitem [{\citenamefont {Nosaka}\ \emph {et~al.}(2018)\citenamefont {Nosaka},
  \citenamefont {Rosa},\ and\ \citenamefont {Yoon}}]{Nosaka:2018iat}%
  \BibitemOpen
  \bibfield  {author} {\bibinfo {author} {\bibfnamefont {T.}~\bibnamefont
  {Nosaka}}, \bibinfo {author} {\bibfnamefont {D.}~\bibnamefont {Rosa}},\ and\
  \bibinfo {author} {\bibfnamefont {J.}~\bibnamefont {Yoon}},\ }\bibfield
  {title} {\bibinfo {title} {{The Thouless time for mass-deformed SYK}},\
  }\href {https://doi.org/10.1007/JHEP09(2018)041} {\bibfield  {journal}
  {\bibinfo  {journal} {Journal of High Energy Physics}\ }\textbf {\bibinfo
  {volume} {2018}},\ \bibinfo {pages} {1} (\bibinfo {year} {2018})}\BibitemShut
  {NoStop}%
\end{thebibliography}%

\end{document}